\documentclass[12pt, draftclsnofoot, onecolumn]{IEEEtran}

\usepackage{ifpdf}
\usepackage{color}
\usepackage{multicol}

\usepackage{cite}

\ifCLASSINFOpdf
  \usepackage[pdftex]{graphicx}

\else

\usepackage[dvips]{graphicx}
\usepackage{setspace}
\fi
\usepackage{ragged2e}
\usepackage{bm}

\usepackage{multicol}
\usepackage{stfloats}

\usepackage[cmex10]{amsmath}

\usepackage{algorithmic}
\usepackage{algorithm}

\usepackage{mathrsfs}

\usepackage{array}

\usepackage{mdwmath}
\usepackage{mdwtab}

\usepackage{eqparbox}

\usepackage{amssymb}

\usepackage[tight,footnotesize]{subfigure}
\begin{document}

\title{Learning-Aided Physical Layer Authentication as an Intelligent Process}

\author{He~Fang,~\IEEEmembership{Student Member, IEEE,}
        Xianbin~Wang,~\IEEEmembership{Fellow, IEEE,} and Lajos~Hanzo,~\IEEEmembership{Fellow, IEEE}

\thanks{H. Fang and X. Wang are with the Department of Electrical and Computer Engineering, The University of Western Ontario, London, ON N6A 5B9, Canada. Email: hfang42@uwo.ca, xianbin.wang@uwo.ca.}

\thanks{L. Hanzo is with the School of Electronics and Computer Science, University of Southampton, Southampton SO17 1BJ, U.K. Email: lh@ecs.soton.ac.uk.}

(Invited Paper)

}

\maketitle

\begin{abstract}
Performance of the existing physical layer authentication schemes could be severely affected by the imperfect estimates and variations of the communication link attributes used.  The commonly adopted static hypothesis testing for physical
layer authentication faces significant challenges in time-varying communication channels due to the changing propagation and interference conditions, which are typically unknown at the design stage.
To circumvent this impediment, we propose an adaptive physical layer authentication scheme based on machine-learning as an intelligent process to learn and utilize the complex and time-varying environment, and hence to improve the reliability and robustness of physical layer authentication.
Explicitly, a physical layer attribute fusion model based on a kernel machine is designed for dealing with multiple attributes without requiring the knowledge of their statistical properties. By modeling the physical layer authentication as a linear system, the proposed technique directly reduces the authentication scope from a combined $N$-dimensional feature space to a single-dimensional (scalar) space, hence leading to reduced authentication complexity. By formulating the learning (training) objective
of the physical layer authentication as a convex problem, an adaptive algorithm based on kernel least-mean-square is then proposed as an intelligent  process to learn and track the variations of multiple attributes, and therefore to enhance the
authentication performance. Both the convergence and the authentication performance of the proposed intelligent authentication process are theoretically analyzed. Our simulations demonstrate that our solution significantly improves the
authentication performance in time-varying environments.
\end{abstract}

\begin{IEEEkeywords}
Intelligent Authentication, Multiple Physical Layer Attributes, Kernel Machine, Adaptive Algorithm
\end{IEEEkeywords}

\IEEEpeerreviewmaketitle

\section{INTRODUCTION}

\IEEEPARstart{D}{ue} to the \emph{open broadcast nature} of radio signal propagation, as well as owing to using \emph{standardized transmission schemes} and \emph{intermittent communications}, wireless communication systems
are extremely vulnerable to interception and spoofing attacks. First of all, the open broadcast nature of the wireless medium facilitates the reception of radio signals by any illegitimate receiver within the coverage of the transmitter \cite{01}. Secondly, the standardized transmission and conventional security schemes of wireless networks make interception and eavesdropping fairly straightforward \cite{50,51}. Moreover, the ``on-off" and sporadic transmissions of low cost wireless devices, especially the significantly growing number of Internet-of-Things (IoT) devices, provide abundant opportunities to adversaries for spoofing attacks. Therefore, the enhancement of authentication schemes is of paramount
 importance for wireless communication systems, especially in the light of the ongoing convergence between the wireless infrastructure and vertical industrial applications enabled by IoT.

\subsection{Comparison of Conventional and Physical Authentication Techniques}
Although digital key-based cryptographic techniques~\cite{35,36,47} have been widely used both for communication security and authentication, they may fall short of the desired performance in many emerging scenarios. One
fundamental weakness of the digital credentials based on conventional cryptography is that detecting compromised security keys cannot be readily achieved, since the inherent physical attributes of communication devices and users are disregarded \cite{01}.
Given the rapidly growing computational capability of low-cost devices, it is becoming more and more feasible to crack the security key from the intercepted signals of standardized and static security protocols.
Furthermore, conventional cryptographic techniques also require appropriate key management procedures to generate, distribute, refresh and revoke digital security keys, which may result in excessive latencies in large-scale networks.
Indeed, this latency may become intolerable for delay-sensitive communications, such as networked control and vehicular communications.  The computational overhead of digital key-based cryptographic methods
 is also particularly undesirable for devices, which have limited battery lifetime and computational capability, such as IoT sensors.

\begin{figure}[htbp]
\centering
\includegraphics[width=16.7cm,height=11cm]{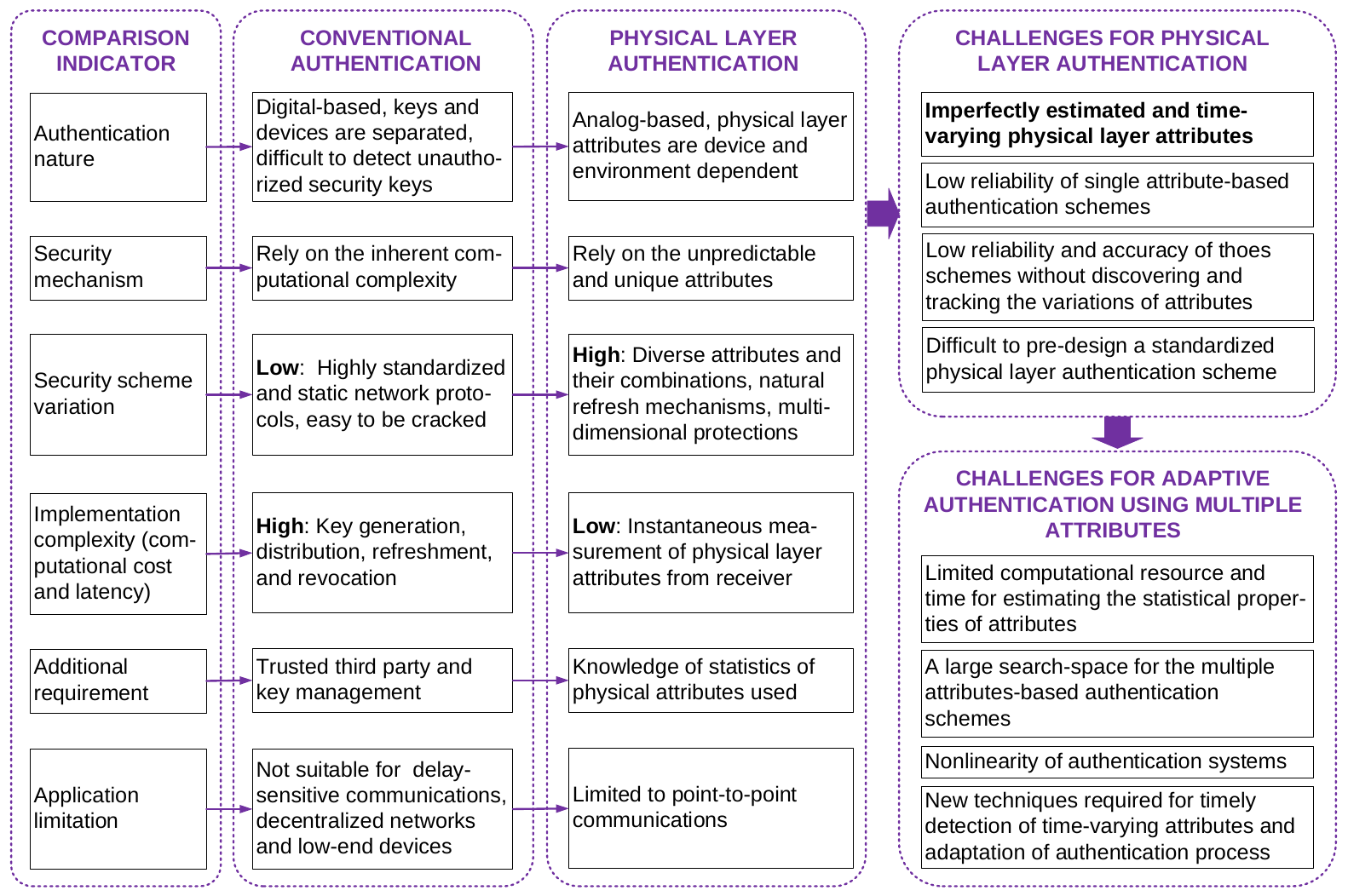}
\caption{\footnotesize{Comparison of conventional and physical authentication techniques.}}
%\label{fig:fig13}
\end{figure}

To overcome these challenges, an alternative approach of authenticating a user (transmitter) is to exploit the physical layer attributes of communication links. Such analog-domain attributes are inherently related to
the unique imperfection of communicating devices and to the corresponding environment, which are hard to impersonate and predict. These physical layer attributes include the channel impulse response (CIR) \cite{40}, received signal
 strength indicator (RSSI) \cite{41}, carrier frequency offset (CFO) \cite{37,38,39}, in-phase-quadrature-phase imbalance (IQI) \cite{39}, and so on, which can also be used to generate  more unique combinations of these attributes
 for authentication. These diverse physical layer attributes and their combinations provide new mechanisms in a multi-dimensional domain for the enhancement of physical layer authentication. Given its obvious advantages including low
 computational requirement, low network overhead and modest energy consumption, physical layer authentication has been widely studied~\cite{02,03,04,05,06,07,08,09,10,11,12,13,14,16,17}. A detailed comparison of conventional and physical
 authentication techniques is given in Fig. 1.

\subsection{Challenges for Physical Layer Authentication}
Despite its many advantages, physical layer authentication also faces several major challenges imposed by the hitherto less well-explored security mechanisms and owing to the analog nature of the link attributes used, as seen in Fig. 1.

\emph{Imperfect estimates} and \emph{variations} of the physical layer attributes are inevitable in practical wireless networks. These constitute challenges for the physical layer authentication, but beneficially,
 they provide unique distinguishing features. Having said that, their adequate estimation often imposes challenges on physical layer authentication, mainly due to time-varying channels, dynamic interference conditions,  mobility of devices, non-symmetrical observations at the transmitter and receiver, as well as owing to the measurement errors, just to name a few.

To elaborate a little further on the challenges, performance of the single-attribute-based physical layer authentication schemes~\cite{02,03,04,05,06,07,08,09,10,11,12,17} remains limited by the imperfect
estimates of the specific attribute used. Moreover, the limited range of the specific attribute distribution may not be sufficiently wide-spread for differentiating the devices all the time. These estimations lead to low-reliability and low-robustness of physical layer
authentication in conjunction with only a single attribute, especially in a hostile time-varying wireless communication environment.

Hence, multiple physical layer attributes may be taken into account for improving the authentication performance~\cite{13,14,16}, since it is more difficult for an adversary to succeed in predicting or imitating all the attributes based on the received signal. On the other hand, when the environment is  time-variant, the performance of physical layer authentication could be severely affected by the unpredicable variations of attributes due to the potential decorrelation of the physical layer attributes
observed at different time instants. Although the variations of attributes provide additional scope for improving the security mechanisms by increasing the uncertainty for the adversaries, at the same time also for the legitimate users operating without discovering and tracking the variations of the attributes.

In a nutshell, the main challenge is that a multiple varying attributes-based authentication scheme is capable of achieving high security in the presence of adversaries, but this increases the grade of challenge imposed on the legitimate users as well.
More importantly, variations of the
physical layer attributes are typically unknown at the design stage and they are hard to predict, thus it is very difficult to pre-design a static  physical layer authentication scheme.
Hence the conception of  an adaptive physical layer authentication scheme  is extremely helpful for improving the performance of physical layer authentication, which can promptly adapt to the time-varying environment. However, designing
near-instantaneously adaptive physical layer authentication based on multiple attributes in rapidly time-varying environments is challenging due to the following reasons:
\begin{itemize}
\item \textbf{C1}. Both the computational resources and the time available for estimating the statistical properties of the physical layer attributes are limited;
\item \textbf{C2}. New authentication schemes based on multiple attributes result in a large search-space, which may lead to both excessive complexity and to non-convex search as well as optimization problems;
\item \textbf{C3}. In practical wireless communication, the typical authentication schemes rely on nonlinear techniques, as exemplified by the binary hypothesis tests of \cite{03,04,05} and by the generalized likelihood ratio test of \cite{02};
\item \textbf{C4}. Timely detection of time-varying physical layer attributes and the adaptation of the physical layer authentication process require sophisticated near-instantaneously adaptive processing techniques.
\end{itemize}

In order to overcome these difficulties, the kernel-based machine learning technique of~\cite{18,19,22,23} is applied for modeling the authentication problem in this paper.
Although the family of parametric learning methods has become mature in the literature~\cite{62,63,69,70,71}, as exemplified by the  linear regression methods of~\cite{62} and the polynomial regression methods of~\cite{63,69}, these parametric techniques usually rely on the assumption of knowing the distribution of samples (i.e. the estimates of the physical layer attributes) together with the specific form of the training function (e.g. linear function or polynomial function). When the assumptions related to the samples' distribution are correct, these parametric methods are more accurate than the non-parametric methods. However, once the assumptions concerning the samples' distribution models become inaccurate, they have a greater chance of failing.
This dramatically limits the employment of parametric learning methods in practical dynamic wireless environments when they face challenge \textbf{C1}, since computing accurate distributions for  multiple physical layer attributes in a complex time-varying environment becomes time-consuming.

In contrast to parametric learning methods, the non-parametric methods are not specified \emph{a priori}, but are determined from the data available.
Hence, the non-parametric methods are more suitable for tracking dynamically time-varying environments without requiring any assumptions concerning the attributes' statistical distributions. Some examples are constituted by the classic k-nearest neighbors~\cite{61} and the decision tree based solutions~\cite{64}. However, these two non-parametric methods have a limited ability to deal with challenges \textbf{C2}-\textbf{C4}. In detail,
it is not easy to determine the most appropriate k-distance in the k-nearest neighbors method. As the decision tree method, the perturbation of the collected data (e.g. by noise) will result in quite a different decision tree, thus leading to quite a different authentication result.

The authors of~\cite{60} proposed a physical layer authentication scheme based on the extreme learning machine concept for improving the spoofing detection accuracy.
The extreme machine learning-based method is basically a 2-layer neural network in which the first layer relies on random parameters, while the second layer is trained by relying on the Moore-Penrose generalized inverse. One of the advantages is its efficiency when compared to the conventional neural network based method invoking back propagation. However, its efficiency critically depends more on the training data set available. Furthermore, this scheme assumes that all the multiple physical layer attributes obey the same statistical distribution
functions, such as the Gaussian distributions, thus their success remains limited in the complex high-dynamic environment considered in this paper.

To overcome these challenges, a promising alternative approach of modeling the authentication process is to track multiple physical layer attributes based on kernel machine learning. As a benefit, the kernel machine~\cite{18,19,22,23} is capable of reducing the dimensionality of the authentication problem based on multiple attributes. It models the authentication problem as a linear system without requiring the knowledge of the attributes' statistical properties. More importantly, the variations of the physical layer attributes as well as of the environment may be tracked (learnt) by the kernel machine learning.
All these compelling benefits motivate us to propose a novel authentication scheme based on the kernel learning as an intelligent process in the face of time-varying wireless communication scenarios to achieve reliable authentication through discovering the complex dynamic environment encountered and through tracking the variations of multiple physical layer attributes.

\subsection{Contributions}

In this paper, we develop an adaptive authentication scheme based on an intelligent machine learning-aided process for discovering  the associated time-varying environment, and thus for improving the physical layer authentication performance. Firstly, a multiple physical layer attribute  fusion model based on the classic kernel
machine is designed for modeling the authentication problem
without requiring the knowledge of those attributes' statistical properties, which corresponds to \textbf{C1} of Section I-B.
As for \textbf{C2} and \textbf{C3}, we cast the authentication problem from a high-dimensional search space to a single-dimensional space by using the classic Gaussian kernel, hence the resultant physical layer authentication can be considered as a linear system. Then an adaptive algorithm is proposed for tracking the variations of the physical layer attributes to achieve a reliable authentication performance, which is a solution for \textbf{C4} of Section I-B.

\textbf{Specifically, the contributions of this paper are summarized as follows:}\\
1) We design a kernel machine-based model for determining the authentication attributes
without requiring the knowledge of their statistical properties, and cast the authentication system from a high-dimensional space to a single-dimensional space.  Then the resultant physical layer authentication process can be considered as a linear system, which is easier to train based on the estimates of the physical layer attributes and on the authentication results observed.
As a result of this transformation, the complexity of our multiple physical layer attribute  fusion model can be dramatically reduced, despite considering a high number of physical layer attributes;\\
2) The learning (training) objective of the physical layer authentication based on kernel machine may be formulated
as a convex problem. We then propose an intelligent authentication process based on kernel least-mean-square for tracking the variations of the physical layer attributes to achieve a reliable authentication performance. By deriving the learning rules for both the system parameters and for the authentication system, the proposed intelligent authentication process becomes capable of adapting to time-varying environments. Therefore, a timely detection of the physical layer attributes and the adjustment of the authentication process can be achieved; \\
3) Our numerical performance and simulations results demonstrate that a larger number of physical layer attributes leads to a more pronounced authentication performance improvement without
 unduly degrading the convergence and training performance. We also demonstrate the superiority of our authentication process over its non-adaptive benchmarker.

The rest of this paper is organized as follows. In Section II, the system model used in this paper is presented. In Section III, we design a multiple physical layer attribute  fusion model based on the kernel machine. An adaptive authentication algorithm is proposed in Section IV, and both the convergence as well as our authentication performance analysis are presented in Section IV. The simulation results are discussed in Section V. Finally, Section VI concludes the paper.

\section{SYSTEM MODEL}

As shown in Fig. 2, we consider a wireless network, where Alice and Bob communicate with each other in the presence of an eavesdropper, explicitly, Eve, who intends to intercept and impersonate Alice, and then to send spoofing signals to obtain illegal advantages. Bob's main objective is to uniquely and unambiguously identify the transmitter by physical layer authentication.
The basic physical layer authentication aims for supporting this pair of legitimate devices by a reciprocal wireless link, while the device-dependent features  can be used as a unique security signature.

\begin{figure}[htbp]
\centering
\includegraphics[width=13.5cm,height=8.2cm]{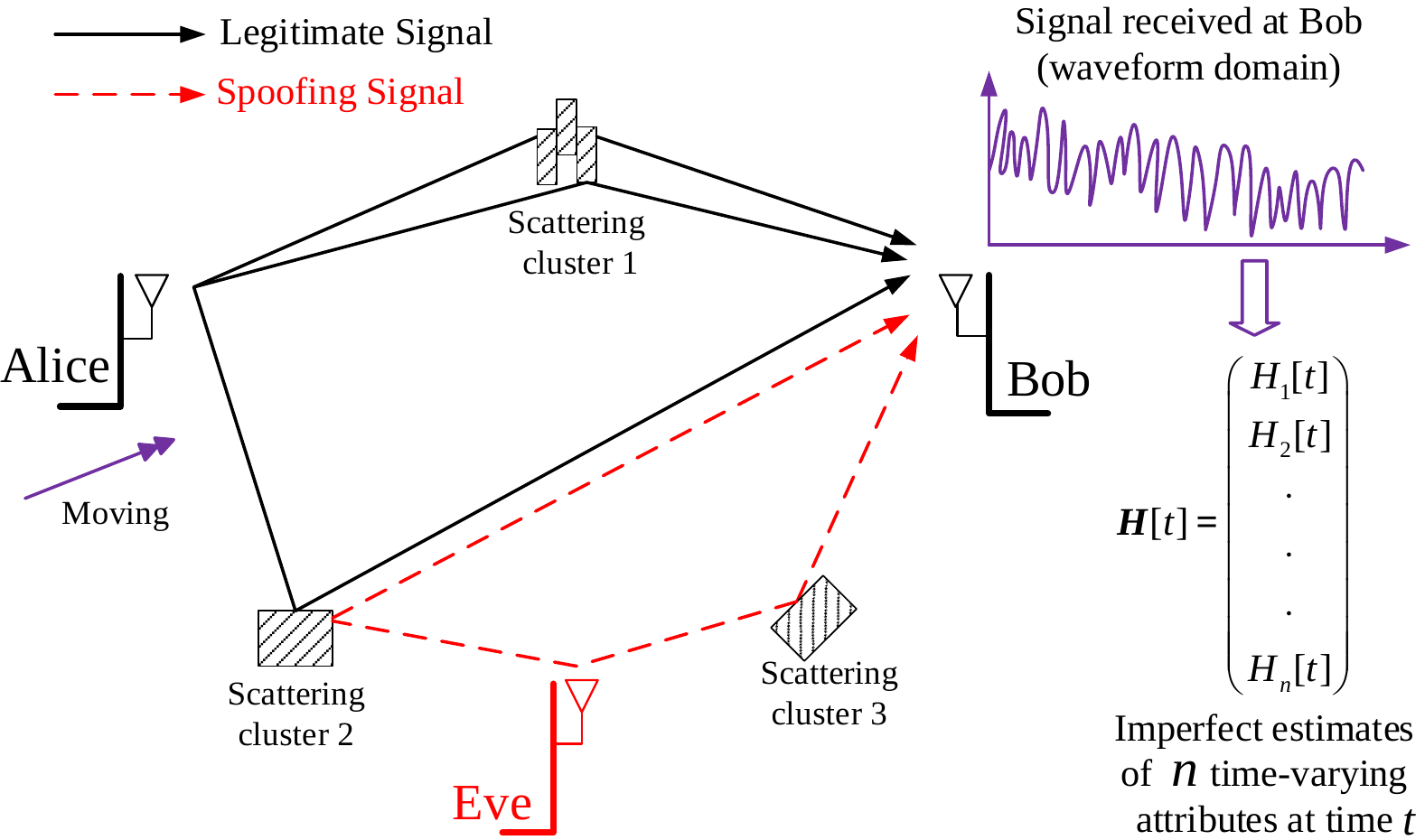}
\caption{\footnotesize{Adversarial system in a wireless network. The transmissions between two legitimate devices (i.e. Alice and Bob) suffer from the spoofing attacks from an attacker, i.e., Eve. Bob should identify the transmitter by using multiple time-varying and imperfectly estimated physical layer attributes.}}
%\label{fig:fig13}
\end{figure}

Again, a combination of multiple attributes can be used for improving the authentication performance, since it is more difficult for Eve to simultaneously infer multiple attributes of a large search-space from a received signal. Naturally, the
various combinations of physical layer attributes provide a high grade of uncertainty for the adversaries and simultaneously improve multi-dimensional protection for the legitimate users.
Let us denote the number of physical layer attributes used for authentication by $N$ and the estimates of multiple physical layer attributes by $\bm{H}=(H_{1},H_{2},...,H_{N})^{{\rm{T}}}$, where ${\rm{T}}$ represents the transposition of a vector. Again, these physical layer attributes may include the channel state information (CSI), carrier frequency offset (CFO), received signal strength indicator (RSSI), round-trip time (RTT), in-phase-quadrature-phase imbalance (IQI), and so on. These unique channel and device features offer security guarantee by physical layer authentication.

Let us continue by stipulating a few important assumptions for the authentication scenario considered in this paper, as follows:\\
\textbf{\emph{Assumption 1.}}
The physical signals transmitted between a pair of legitimate devices rapidly become decorrelated in space, time and frequency. This implies that it is hard for the attacker to observe and predict the channel state between legitimate devices, if the attacker is at a third location, which is further than a wavelength away from Alice and Bob;\\
\textbf{\emph{Assumption 2.}} Both the wireless channels and the interference are time-varying, the devices are moving, and hence the wireless environment is dynamically changing. These all lead to unpredictable variations of the physical layer attributes;\\
\textbf{\emph{Assumption 3.}} The estimates of the physical layer attributes are imperfect, because
the legitimate devices roaming in different locations also suffer from different interferences, a dynamic propagation environment, different estimation errors, and so on.\\
These assumptions characterize a practical scenario, but naturally, it is more difficult for us to deal with these imperfectly estimated time-varying  physical layer attributes.

The physical layer authentication comprises two phases, as described below.\\
 \emph{Phase I:} Alice broadcasts one or more messages to Bob at time $t$. From the received signal, Bob infers an imperfect estimate of the multiple attributes
 \begin{eqnarray}
\bm{H}_{A}^{I}[t]=(H_{A1}^{I}[t], H_{A2}^{I}[t],...,H_{AN}^{I}[t])^{{\rm{T}}},
\end{eqnarray}
which are associated with Alice. At the same time, Eve overhears the transmission.\\
\emph{Phase II:} either Alice or Eve transmits a message to Bob at time $t+\tau$. Then Bob obtains another imperfect estimate
\begin{eqnarray}
\bm{H}^{II}[t+\tau]=(H^{II}_{1}[t+\tau],H^{II}_{2}[t+\tau],...,H^{II}_{N}[t+\tau])^{{\rm{T}}},
\end{eqnarray}
where $\tau$ represents the time interval between the two phases.

Bob should compare the estimate $\bm{H}^{II}[t+\tau]$ to the previous estimate $\bm{H}_{A}^{I}[t]$. If these two estimates are likely to be originated from the same channel realization and the same imperfect hardware, then the message is deemed to be coming from Alice. \\
\textbf{Remark 1}. As we mentioned in the assumptions,
the physical layer attributes are time-variant and imperfectly estimated. The objective of this paper is to propose an intelligent authentication process relying on these physical layer attributes.
The process proposed
aims for achieving reliable and robust authentication through discovering and learning the complex operating environment, in the face of limited
computational resources (see \textbf{C1});
our new authentication schemes based on multiple attributes result in a higher-dimensional search space (\textbf{C2});
the authentication schemes usually rely on nonlinear processing (\textbf{C3});
the prompt detection of the time-varying physical layer attributes and the ensuing adjustment of the physical layer authentication require new sophisticated adaptive processing techniques (\textbf{C4}).

Let us now conceive an intelligent adaptive function $\mathcal{F}(\cdot)$, which is used for fusing $N$ independent physical layer attributes. Then the authentication process can be formulated as a binary hypothesis test relying on a threshold $\nu>0$
\begin{eqnarray}
\begin{cases}
\Phi_{0}:&~ |\mathcal{F}(\bm{H}_{A}^{I}-\bm{H}^{II})|\leq\nu;\\
\Phi_{1}:&~ |\mathcal{F}(\bm{H}_{A}^{I}-\bm{H}^{II})|>\nu,
\end{cases}
\end{eqnarray}
where $\Phi_{0}$ indicates that the signal is from Alice, while $\Phi_{1}$ indicates that it is from Eve. Due to the variations and imperfect estimates of the physical layer attributes between Alice and Bob, we may encounter both
false alarms and misdetections. Therefore, the parameters in $\mathcal{F}(\cdot)$ should be promptly updated to achieve low false alarm rate and misdetection rate in a time-varying environment.

\section{KERNEL MACHINE-BASED MULTIPLE PHYSICAL LAYER ATTRIBUTE FUSION}
In order to improve the performance of the authentication schemes in time-varying environments using multiple physical layer attributes, which are \emph{imperfectly estimated} and \emph{time-varying}, we propose  a kernel machine-based model for fusing multiple physical layer attributes without requiring the knowledge of their statical properties  in the spirit of \textbf{C1} of Section I-B.
 Then,
the dimension of the search-space is reduced from $N$ to 1 with the aid of our kernel machine-based physical layer attribute  fusion model and
our authentication problem can be modeled by a linear system as detailed in this section (corresponding to \textbf{C2} and \textbf{C3} of Section I-B).
Therefore,
the trade-off between the authentication
false alarm and misdetection can be improved.

\begin{figure}[htbp]
\centering
\includegraphics[width=8.8cm,height=2.5cm]{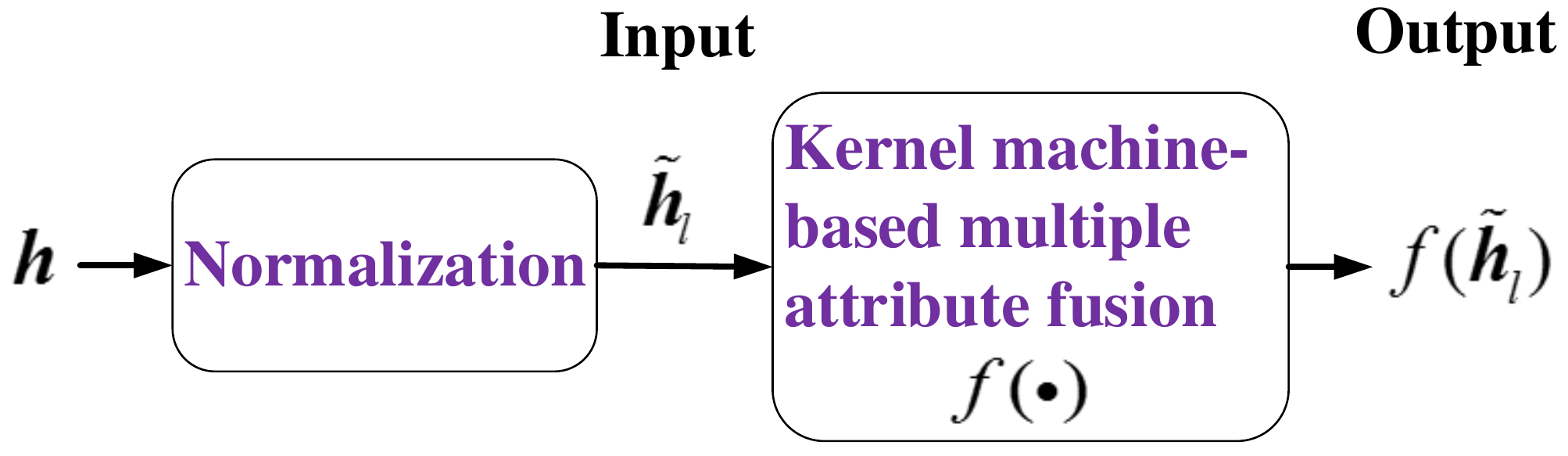}
\caption{\footnotesize{Kernel machine-based multiple physical layer attribute fusion.}}
%\label{fig:fig13}
\end{figure}

In the kernel machine based multiple attribute fusion, Bob will obtain an estimate $\bm{H}^{II}[t+\tau]$ of (2) at time $t+\tau$. Then, Bob will compare the estimate $\bm{H}^{II}[t+\tau]$ to the previous estimate at time $t$, namely for $\bm{H}^{I}_{A}[t]$ of (1). The difference between these two estimates is denoted as $\bm{h}=(h_{1},h_{2},...,h_{N})^{{\rm{T}}}$, where each $h_{n}\in [a_{n},b_{n}]$ is formulated as
\begin{eqnarray}
h_{n}=H_{An}^{I}[t]-H_{n}^{II}[t+\tau],~n=1,...,N,
\end{eqnarray}
with $N$ being the number of physical layer attributes used.

Since the different attributes exhibit quite different ranges and have different units, the normalization (see Fig. 3) is required for scaling the attributes having different ranges to the same range for the ease of analysis and for the design of the kernel machine-based fusion. In the following, we normalize the attributes having ranges $[a_{n},b_{n}], n=1,...,N,$ to $[-1,1]$ by invoking
\begin{eqnarray}
\widetilde{h}_{n}=\frac{2}{b_{n}-a_{n}}(h_{n}-\frac{a_{n}+b_{n}}{2}),~n=1,...,N.
\end{eqnarray}

It can be observed from (4) and  (5) that these two equations are only used for normalizing the estimates of the attributes to the range of $[-1,1]$, so that the rather diverse multiple physical layer attributes can be processed in the same range.  In practical systems, we only have to know the approximate variation ranges of the attributes, which is reasonable because we can always have \emph{a priori} knowledge about the communication systems and environments before designing the authentication system. For example, the CFO variation range was measured to be $[-78.125,78.125)$ kHz according to \cite{06}, while the RSSI range depends on the pathloss. If $a_{n}$ and $b_{n}$ are excessive, the estimates of attribute $n$ will be scaled to a smaller real range instead of the nominal range of $[-1,1]$, for example, to $[-0.5,0.5]$. If the real range is not too small, for example, not hundreds of times smaller than the nominal range of $[-1,1]$, this kind of scaling problems can be readily overcome  and will not unduly degrade the authentication performance. We will demonstrate this result and discuss a bit further in Fig. 12 of Section V.

Let us assume that a set of observations $(\bm{\widetilde{h}}_{l},\widehat{y}_{l})_{l=1}^{L}\in [-1,1]^{N}\times \{0,1\}$ is given, which is used for training the authentication process, where
$\bm{\widetilde{h}}_{l}=(\widetilde{h}_{1l},\widetilde{h}_{2l},...,\widetilde{h}_{Nl})^{{\rm{T}}}$ is the $l$th estimate after the normalization, with  each element $\widetilde{h}_{nl}$ defined in (5), and
\begin{eqnarray}
\widehat{y}_{l}=
\begin{cases}
1 & \Phi_{0}\\
0 & \Phi_{1}
\end{cases}.
\end{eqnarray}
 As shown in Fig. 3, the normalized estimates $\bm{\widetilde{h}}_{l},l=1,2,...,L,$ are considered as the inputs of the kernel machine and $f(\bm{\widetilde{h}}_{l})$
represent the outputs of the kernel machine with the corresponding inputs given by $\bm{\widetilde{h}}_{l}\in [-1,1]^{N},l=1,2,...,L$. Note that for the
legitimate users, the training data of a legitimate communication session
is relatively straightforward to obtain \cite{65}.

Our task is then to infer the underlying mapping function $\widehat{y}_{l}=f(\bm{\widetilde{h}}_{l})$ from the training data set (the samples) received $(\bm{\widetilde{h}}_{l},\widehat{y}_{l})_{l=1}^{L}\in [-1,1]^{N}\times \{0,1\}$. In other words, the task in this section is to represent the authentication system $\widehat{y}_{l}=f(\bm{\widetilde{h}}_{l})$ and to model the relationship between the estimates of the multiple attributes and the corresponding authentication results. After this, we can verify whether a transmitter is that of Alice or of Eve once a new normalized estimate $\bm{\overline{h}}=(\overline{h}_{1},\overline{h}_{2},...,\overline{h}_{N})^{{\rm{T}}}$ has been obtained. For example, in a continuous authentication session as defined in~\cite{17}, once a transmitter accesses the system again, Bob can infer the estimate of this transmitter's physical attributes one more time and then determine its normalized estimate through (5). This normalized estimate may be different from the previous normalized estimates $\bm{\widetilde{h}}_{l},l=1,2,...,L,$ because of the time-varying environment or channels, which will be treated as the new normalized estimate of the attributes. The authentication is then  carried out by using the new normalized estimate to improve the security. \\
\textbf{Definition 1}~\cite{18}: A \emph{Mercer kernel} is a continuous, symmetric, positive-definite function $\kappa: [-1,1]^{N}\times [-1,1]^{N}\rightarrow \mathcal{R}$.

The classic Gaussian kernel function of~\cite{18,19,22,23} is adopted, which has an excellent modelling capability and it is also numerically stable. The Gaussian kernel function is given by
\begin{eqnarray}
\kappa(\bm{\widetilde{h}}_{l},\bm{\overline{h}})=\exp (\frac{-\|\bm{\widetilde{h}}_{l}-\bm{\overline{h}}\|^{2}}{2\sigma^{2}}),
\end{eqnarray}
where $\sigma$ is the kernel width. The kernel width $\sigma$ should be chosen by the users. The popular methods of selecting a suitable kernel width include the cross-validation, nearest neighbor, penalizing function and plug-in based methods of~\cite{28}. The Gaussian kernel function of (7) characterizes a similarity between the observed inputs $\bm{\widetilde{h}}_{l}$ and the new normalized estimate $\bm{\overline{h}}$.

The kernel machine projects the $N$-dimensional input vector $\bm{\overline{h}}\in [-1,1]^{N}$ into a potentially infinite-dimensional feature space $\mathcal{H}$ through a mapping $\varphi: [-1,1]^{N}\rightarrow \mathcal{H}$. Note that the transformation from the input space into the feature space is nonlinear, and the dimensionality of the feature space is high enough. Then the authentication system can be formulated as
\begin{eqnarray}
f(\bm{\overline{h}})=\bm{w}^{{\rm{T}}}\varphi (\bm{\overline{h}}),
\end{eqnarray}
where $\bm{w}$ is the weight vector in the feature space $\mathcal{H}$.

Given the Gaussian kernel function of (7), we can obtain a so-called reproducing kernel Hilbert space defined as follows:\\
\textbf{Definition 2}~\cite{24}: Consider a Hilbert space $\mathcal{H}$ constituted by the functions $g: [-1,1]^{N}\rightarrow \mathcal{R}$. It is called a \emph{reproducing kernel Hilbert space} if there exists a kernel $\kappa: [-1,1]^{N}\times [-1,1]^{N}\rightarrow \mathcal{R}$ having the following properties:
\begin{itemize}
\item $\kappa$ has the reproducing property
\begin{eqnarray}
\langle g, \kappa(\bm{\widetilde{h}}_{l},\cdot)\rangle=g(\bm{\widetilde{h}}_{l}),~ {\rm{for~all}}~g\in \mathcal{H},
\end{eqnarray}
\item $\kappa$ spans $\mathcal{H}$, i.e. $\mathcal{H}$ is the completion of a vector space spanned by $\kappa(\bm{\widetilde{h}}_{l},\cdot)$ for $\bm{\widetilde{h}}_{l}\in [-1,1]^{N}$.
\end{itemize}

Therefore, the feature space $\mathcal{H}$ spanned by the Gaussian kernel function of (7) is a reproducing kernel Hilbert space. We now impose the Representer theorem of~\cite{24,25}, which is given by: \\
\textbf{Representer Theorem}~\cite{24,25}: Let $\Omega: [0,\infty)\rightarrow \mathcal{R}$ be a strictly monotonically increasing function, $\mathcal{U}$ be a nonempty set, $c: (\mathcal{U}\times \mathcal{R}^{2})^{L}\rightarrow \mathcal{R} \bigcup \{\infty\}$ be an arbitrary risk function, and $\mathcal{H}$ be the reproducing kernel Hilbert space associated with a kernel $\kappa(\bm{\widetilde{h}}_{l},\cdot)$. Then each minimizer $g\in \mathcal{H}$ of the regularized risk function
\begin{eqnarray}
c\{[\bm{\widetilde{h}}_{1},\widehat{y}_{1},g(\bm{\widetilde{h}}_{1})],...,[\bm{\widetilde{h}}_{L},\widehat{y}_{L},g(\bm{\widetilde{h}}_{L})]\}+\Omega (\| g \|_{\mathcal{H}}),
\end{eqnarray}
lends itself to a representation of the form
\begin{eqnarray}
g(\cdot)=\sum_{l=1}^{L}\alpha_{l}\kappa(\bm{\widetilde{h}}_{l},\cdot).
\end{eqnarray}

Therefore, given the Gaussian kernel function of (7), according to the Representer Theorem, the optimal authentication system expression can be formulated as
\begin{eqnarray}
f(\bm{\overline{h}})=\sum_{l=1}^{L}\alpha_{l}\kappa(\bm{\widetilde{h}}_{l},\bm{\overline{h}}).
\end{eqnarray}
An important relationship between (8) and (12) is
\begin{eqnarray}
\kappa(\bm{\widetilde{h}}_{l}, \bm{\overline{h}})=\varphi (\bm{\widetilde{h}}_{l})^{{\rm{T}}}\varphi (\bm{\overline{h}}).
\end{eqnarray}
\textbf{Remark 2}. We can observe from the kernel function of (7) and from the optimal authentication system expression of (12) that the physical layer attributes are fused without any specific knowledge of their statical properties,
which corresponds to \textbf{C1} of Section I-B. As for  \textbf{C2}, the search-space is transformed from being $N$-dimensional to single-dimensional by our multiple physical layer attribute fusion model. \\
\textbf{Remark 3}. In practical wireless networks, the authentication systems are usually nonlinear. For example, the binary hypothesis test of \cite{03,04,05} and the generalized likelihood ratio test of \cite{02} (see \textbf{C3} in Section I-B)
can be imposed. By contrast, according to the proposed kernel machine-based physical layer attribute fusion model of (12), the authentication system is formulated as a linear system, since the  expression of (12)  relies on the linear weights $\alpha_{l},~l=1,2,...,L$.

As discussed above, the estimates of the multiple physical layer attributes $\bm{H}$ are time-variant, which may lead to a low authentication performance without adaptation. Therefore, in the next section, we focus our attention on proposing an adaptive algorithm for promptly adjusting the authentication system and for updating the parameters in (12), i.e. $\alpha_{l},~l=1,2,...,L$, through discovering and learning the complex dynamic environment encountered, which is the solution of \textbf{C4} in Section I-B.

\section{ADAPTIVE AUTHENTICATION AS AN INTELLIGENT PROCESS}
In this section, a learning procedure is proposed for adaptive authentication based on the kernel least-mean-square for promptly updating the parameters. This authentication process is based on learning from the observed samples
$(\bm{\widetilde{h}}_{l},\widehat{y}_{l})_{l=1}^{L}\in [-1,1]^{N}\times \{0,1\}$. Explicitly, the proposed learning procedure can be viewed as an intelligent process of learning the time-varying environment for updating the system parameters $\alpha_{l},~l=1,2,...,L$,
to achieve reliable and robust authentication.

\subsection{Adaptive Authentication Algorithm}
Given the samples $(\bm{\widetilde{h}}_{l},\widehat{y}_{l})_{l=1}^{L}\in [-1,1]^{N}\times \{0,1\}$ observed, we transform the $N$-dimensional input vector $\bm{\widetilde{h}}_{l} \in [-1,1]^{N}$ into a kernel Hilbert space $\mathcal{H}$ through a mapping $\varphi: [-1,1]^{N}\rightarrow \mathcal{H}$ according to (8). Therefore, we obtain a pair of sample sequences $\{\varphi (\bm{\widetilde{h}}_{1}), \varphi (\bm{\widetilde{h}}_{2}),...\}$ and $\{\widehat{y}_{1},\widehat{y}_{2},... \}$. The weight vector $\bm{w}$ in (8) at iteration $l$ should be updated for minimizing the cost function as follows
\begin{eqnarray}
\min_{\bm{w}}\sum _{k=1}^{l}[\widehat{y}_{k}-\bm{w}^{{\rm{T}}}\varphi (\bm{\widetilde{h}}_{k})]^{2}.
\end{eqnarray}
\textbf{Remark 4}. We can observe from (14) that the learning (training) objective of the adaptive authentication process is formulated as a convex optimization problem.

Then the learning procedures conceived for updating the weight vector $\bm{\alpha}$ and the authentication system of (12) are given by the following theorems:\\
\textbf{Theorem 1}: The learning rule conceived for updating the weight vector $\bm{\alpha}[l]$ in our multiple physical layer attribute  fusion model at iteration $l$ can be expressed as
\begin{eqnarray}
\bm{\alpha}[l]
  =\mu \times ( e[1],e[2],...,e[l])^{{\rm{T}}},
\end{eqnarray}
where $\mu$ represents a step-size parameter. Furthermore, $e[l]$ is the prediction error computed as the difference between the desired observation of the transmitter and its prediction relying on the authentication system parameters
$\bm{\alpha}[l-1]$, which is expressed as
\begin{eqnarray}
e[l]=\widehat{y}_{l}-f(\bm{\widetilde{h}}_{l})[l-1],
\end{eqnarray}
where we have
\begin{eqnarray}
f(\bm{\widetilde{h}}_{l})[l-1]=\sum_{i=1}^{l-1}\alpha_{i}[l-1]\kappa(\bm{\widetilde{h}}_{i},\bm{\widetilde{h}}_{l}).
\end{eqnarray}
\textbf{Proof}: See Appendix A.\\
\textbf{Theorem 2}:
The learning rule conceived for adjusting the authentication system at iteration $l$ is given by
\begin{eqnarray}
f(\bm{\overline{h}})[l]=f(\bm{\overline{h}})[l-1]+ \mu e[l]\kappa(\bm{\widetilde{h}}_{l},\bm{\overline{h}}).
\end{eqnarray}
\textbf{Proof}: See Appendix B.

In conclusion, according Theorem 1 and Theorem 2, our intelligent authentication process based on the kernel least-mean-square is summarized at a glance in Algorithm 1.
\begin{algorithm}
    \caption{Intelligent authentication process}
    \label{alg:Framwork}
    \begin{algorithmic}
        \STATE \textbf{1.~Initialization:}
        \STATE ~~~$f[0]=0$: initial value of authentication system
         \STATE ~~~$e[0]=0$: initial value of prediction error
         \STATE ~~~$\alpha[0]=0$: initial value of system parameter $\bm{\alpha}$
         \STATE ~~~$\mu$: step-size parameter of learning
         \STATE ~~~$\sigma$: kernel width
         \STATE ~~~$\bm{\widetilde{h}}_{1}$: initial input, i.e. the normalized estimate of physical layer attributes
         \STATE ~~~$\mathcal{C}=\{\bm{\widetilde{h}}_{1}\}$: initial set of input
          \STATE ~~~$\widehat{y}_{1}$: initial observation of the transmitter with the corresponding normalized estimate $\bm{\widetilde{h}}_{1}$
          \STATE \textbf{2.~Iteration:}
         \STATE ~~~\textbf{2.1} \textbf{while} samples $(\bm{\widetilde{h}}_{l},\widehat{y}_{l})_{l=1}^{L}\in [-1,1]^{N}\times \{0,1\}$ available \textbf{do}
         \STATE ~~~\textbf{2.2} ~~~obtain the output of authentication system $f[l-1]$ at iteration $l-1$ via (12);
         \STATE ~~~\textbf{2.3} ~~~calculate the prediction error $e[l]$ via (16);
        \STATE ~~~\textbf{2.4} ~~~update weight vector $\bm{\alpha}[l]$ through (15);
           \STATE ~~~\textbf{2.5} ~~~adjust the authentication system $f[l]$ via (18);
        \STATE ~~~\textbf{2.6} ~~~update the input set as $\mathcal{C}=\mathcal{C}+\{\bm{\widetilde{h}}_{l}\}$;
        \STATE ~~~\textbf{2.7} ~~~$l=l+1$;
            \STATE ~~~\textbf{2.8} \textbf{end}
    \end{algorithmic}
\end{algorithm}\\
\textbf{Remark 5}. In conclusion,  the search space is transformed from being $N$-dimensional to single-dimensional (see Remark 2), the authentication is modelled as a linear system (see Remark 3), and the learning
(training) objective of the authentication is formulated as a convex problem (see Remark 4). Therefore, it dramatically reduces the complexity of our physical layer authentication technique relying on multiple attributes.
We can also observe from Algorithm 1 that there is only one `while' loop in step 2, so the execution-time is on the order of $O(L)$, which makes Algorithm 1 an attractive solution.
In the next subsection, we will discuss the selection of the step-size parameter $\mu$, which affects the convergence of our authentication process.

\subsection{Convergence of the Proposed Authentication Process}
The step-size parameter directly affects the convergence of our authentication process, since increasing the step-size of learning will reduce the convergence time but may in fact lead to divergence. Therefore, the step-size parameter $\mu$ should be carefully decided.\\
\textbf{Theorem 3}:
The proposed intelligent authentication process (see Algorithm 1) converges to a steady-state value, if the step-size parameter of learning $\mu$ satisfies
\begin{eqnarray}
0<\mu<\frac{L}{\sum_{l=1}^{L}\kappa(\bm{\widetilde{h}}_{l},\bm{\widetilde{h}}_{l})}.
\end{eqnarray}
\textbf{Proof}. See Appendix C. \\
\textbf{Remark 6}. Theorem 3 gives the upper bound of the step-size parameter $\mu$ in Algorithm 1, so that the proposed intelligent authentication process converges to a steady state.

\subsection{Authentication Performance Analysis}

Mathematically, the false alarm rate and the misdetection rate can be formulated based on the hypothesis test of (3), respectively, as
\begin{eqnarray}
P_{FA}=P(|\mathcal{F}(\bm{H}_{A}^{I}-\bm{H}^{II})|>\nu \mid \Phi_{0})
\end{eqnarray}
and
\begin{eqnarray}
P_{MD}=P(|\mathcal{F}(\bm{H}_{A}^{I}-\bm{H}^{II})|\leq\nu \mid \Phi_{1}),
\end{eqnarray}
where $\mathcal{F}$ represents our multiple physical layer attribute fusion model.

According to the proposed authentication system of (12), the false alarm rate and misdetection rate at instant $L$ can be rewritten, respectively, as
\begin{eqnarray}
P_{FA}=P(|\sum_{l=1}^{L-1}\alpha_{l}\kappa(\bm{\widetilde{h}}_{l},\bm{\widetilde{h}}_{L})|\leq\nu \mid \Phi_{0})
\end{eqnarray}
and
\begin{eqnarray}
P_{MD}=P(|1- \sum_{l=1}^{L-1}\alpha_{l}\kappa(\bm{\widetilde{h}}_{l},\bm{\widetilde{h}}_{L})|\leq\nu \mid \Phi_{1}),
\end{eqnarray}
where $\nu\in [0,1)$.

In the face of the imperfect estimates of time-varying physical layer attributes, we divide them into two parts:
the time-varying part $\bm{\overline{H}}$ that is the real value of physical layer attributes, and part $\bigtriangleup \bm{H}$ that is the bias of estimated attributes. Then the estimates $\bm{H}_{A}^{I}[l-\tau_{l}]$ and $\bm{H}^{II}[l]$ can be written, respectively, as
\begin{eqnarray}
\bm{H}_{A}^{I}[l-\tau_{l}]=\bm{\overline{H}}_{A}^{I}[l-\tau_{l}]+\bigtriangleup \bm{H}_{A}^{I}[l-\tau_{l}]
\end{eqnarray}
and
\begin{eqnarray}
 %\nonumber to remove numbering (before each equation)
\bm{H}^{II}[l]=\bm{\overline{H}}^{II}[l]+\bigtriangleup \bm{H}^{II}[l],
\end{eqnarray}
where $\tau_{l}$ is the time interval between Phase I and Phase II of the physical layer authentication at iteration $l$.
Furthermore, $\bm{\upsilon}(\tau_{l})=(\upsilon_{1l},\upsilon_{2l},...,\upsilon_{Nl})^{{\rm{T}}}$ represents the variations of part $\bm{\overline{H}}_{A}^{I}$ during the time interval $\tau_{l}$, which can be expressed as
\begin{eqnarray}
 %\nonumber to remove numbering (before each equation)
\bm{\upsilon}(\tau_{l})=\bm{\overline{H}}_{A}^{II}[l]-\bm{\overline{H}}_{A}^{I}[l-\tau_{l}].
\end{eqnarray}

%Assuming part $\bigtriangleup \bm{H}$ of the multiple attributes obey zero-mean complex Gaussian distribution with variance $\bm{\varrho}^{2}=(\varrho_{1}^{2},\varrho_{2}^{2},...,\varrho_{n}^{2})^{{\rm{T}}}$. %Besides, we assume the period between Phase I and Phase II of the authentication $\tau$ is the same in every sample.
Given the distributions of part $\bigtriangleup \bm{H}$ of the multiple physical layer attributes,  we can calculate the false alarm rate and misdetection rate of our scheme as
the following theorems. Note that our intelligent authentication process does not need the knowledge of their statistical properties in the training process. \\
\textbf{Theorem 4}: The false alarm rate expression of our intelligent authentication process at iteration $L$ is given by
\begin{eqnarray}
P_{FA}=F_{Y_{1}}\ast F_{Y_{2}}\ast\cdots\ast F_{Y_{L-1}}(\nu)
-F_{Y_{1}}\ast F_{Y_{2}}\ast\cdots\ast F_{Y_{L-1}}(-\nu),
\end{eqnarray}
where $Y_{l}=\alpha_{l}\exp(-\sum_{i=1}^{N}(\widetilde{h}_{il}-\widetilde{h}_{iL}^{\Phi_{0}})^{2}/2\sigma^{2})$, $l=1,2,...,L-1$, $\widetilde{h}_{iL}^{\Phi_{0}}$ is shown in (42), $F$ represents the cumulative distribution function, and $\ast$ represents the convolution.\\
\textbf{Proof}: See Appendix D.\\
\textbf{Theorem 5}: The misdetection rate expression of our intelligent authentication process at iteration $L$ is expressed as
\begin{eqnarray}
P_{MD}=F_{Z_{1}}\ast F_{Z_{2}}\ast\cdots\ast F_{Z_{L-1}}(\nu+1)
-F_{Z_{1}}\ast F_{Z_{2}}\ast \cdots\ast F_{Z_{L-1}}(1-\nu),
\end{eqnarray}
where $Z_{l}=\alpha_{l}\exp(-\sum_{i=1}^{N}(\widetilde{h}_{il}-\widetilde{h}_{iL}^{\Phi_{1}})^{2}/2\sigma^{2})$, $l=1,2,...,L-1$, and $\widetilde{h}_{iL}^{\Phi_{1}}$ is shown in (44). \\
\textbf{Proof}: See Appendix E.\\
\textbf{Remark 7}. We can observe from Theorem 4 and Theorem 5 that the false alarm rate and misdetection  rate of  our intelligent authentication process depend on both the number of physical layer attributes $N$ and on the variations of the attributes $\bm{\upsilon}$. Our intelligent authentication process tracks the variations of the attributes $\bm{\upsilon}$ and promptly adjusts the authentication system, so that a compelling false alarm rate vs. misdetection rate trade-off is achieved.

\section{NUMERICAL PERFORMANCE AND SIMULATION RESULTS}

In order to evaluate the performance of our intelligent authentication process, we provide both numerical and simulation results in this section.
Firstly, we implement our authentication process using some specific physical layer attributes, and characterize the convergence of Algorithm 1. Then its false alarm rate vs. the misdetection rate trade-off is studied. Finally, we demonstrate the superiority of our authentication process over its non-adaptive benchmarker.

First of all, three physical layer attributes, namely the  carrier frequency offset (CFO),  channel impulse response (CIR), and received signal strength indicator (RSSI) are considered in our simulations to confirm the viability of our intelligent authentication process. Specifically, a communication system having a measurement center frequency of 2.5 GHz, measurement bandwidth of 10 MHz, coherence bandwidth of 0.99, normalized time correlation of 0.99 and sampling time of 50 ms is considered. The transmitted signal is passed through a randomly generated 12-tap multipath fading channel having an exponential power delay profile.  We assume that the relative velocity between Alice and Bob is around 20 km/h, and the initial distance between Alice and Bob is 5 m. Then the CFO of an individual transmitter can be approximated as a zero-mean Gaussian variable \cite{06,56}, and the standard deviation of the CFO variation is $\triangle_{\rm{CFO}}\approx 2.35\times 10^{-7}$. The CFO estimation range is $[-78.125,78.125)$ kHz \cite{06}. Furthermore, according to \cite{02},  an autoregressive model of order 1 (AR-1) is used for characterizing the temporal amplitude fluctuation ${\rm{Amp}}_{k}[t]$ of the $k$th-tap in our multipath fading channel.
Therefore, the variation of the CIR can be modelled by $\upsilon_{{\rm{CIR}}}=\sum _{k=1}^{12}{\rm{Amp}}_{k}[t]\exp(-j4.99\pi k)$, and the per-tone signal-to-noise ratio (SNR) is in the channel measurements range of $[-12.8, 14.2)$ dB with a median value of 6.4 dB, if the transmit power is 10 mW \cite{02}.
Finally, according to \cite{67}, the RSSI can be modeled as $PL[\rm{dB}]=75+36.1\log(d/10)$, where $PL$ is the path loss, and $d$ represents the direct transmission distance between the transmitter and Bob.
The direct transmission distance between the transmitter and Bob is assumed to be in the range of $[0,100]$ m.

\begin{figure}[htbp]
\centering
\includegraphics[width=12cm,height=9cm]{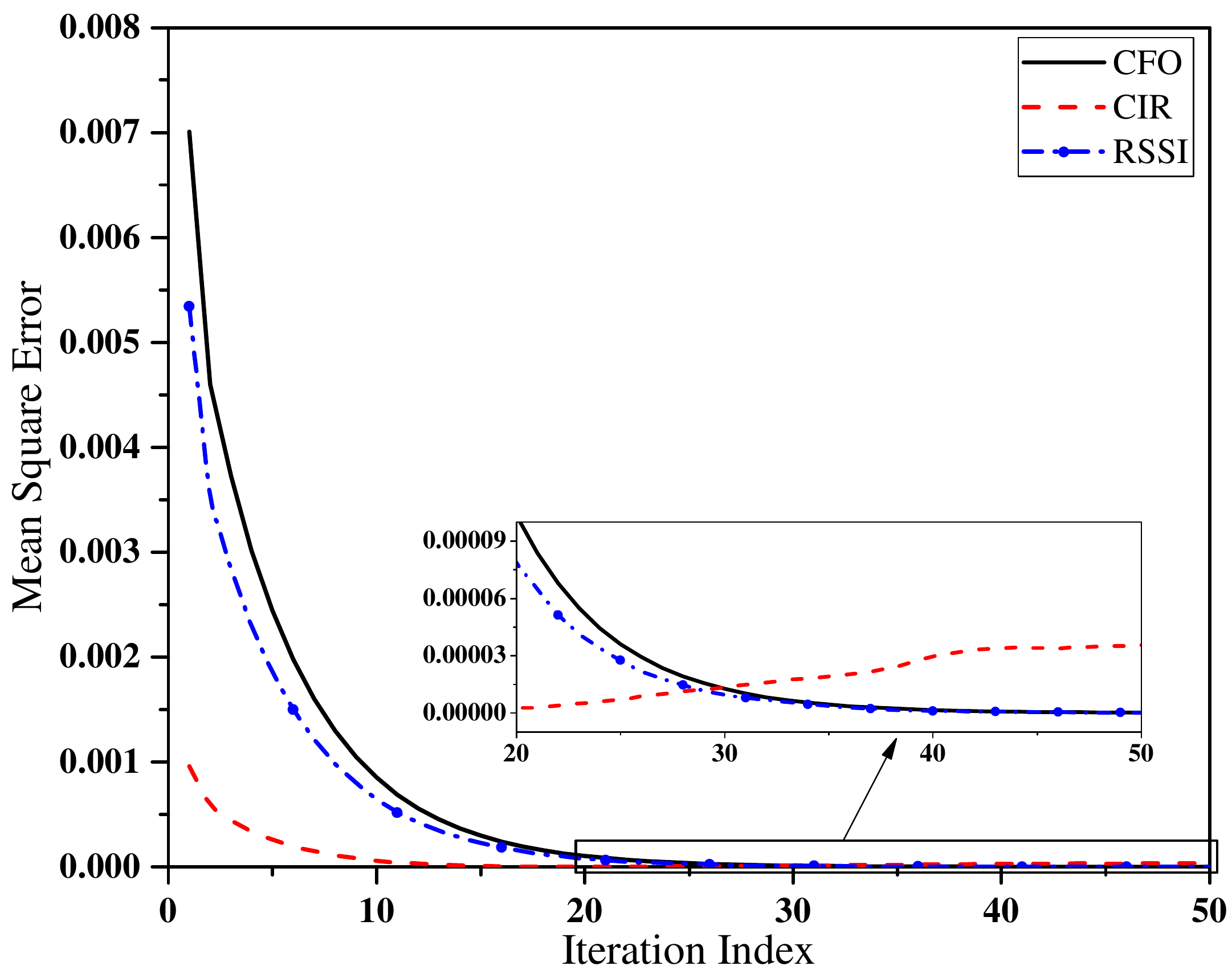}
\caption{\footnotesize{Training performance of our intelligent authentication process (Algorithm 1) relying on the CFO, CIR and RSSI triplet.}}
%\label{fig:fig13}
\end{figure}

Given 300 samples of the CFO, CIR and RSSI of Alice, i.e. $(\bm{\widetilde{h}}_{l},\widehat{y}_{l})_{l=1}^{300}\in [-1,1]^{3}\times \{0,1\}$, where $\widehat{y}_{l}=1$,
 Fig. 4 shows the training performance of our intelligent authentication process (Algorithm 1) relying on the CFO, CIR and RSSI triplet.
The step-size parameter of Algorithm 1 is set to $\mu=0.1$.
We can observe from Fig. 4 that the mean square errors $ E[\| e[l]\|^{2}]$ of all the strategies
are significantly reduced, as the iteration index increases from 0 to 50. Furthermore, the mean square error $ E[\| e[l]\|^{2}]$ of each strategy reaches its steady-state value after 30 iterations.
We can also observe from Fig. 4 that the CIR estimation performs better than both the CFO and RSSI estimation in the training performance at the beginning, but its training performance becomes the worst after 30 iterations. The reason for this trend is that the deviation of CIR estimation is lower than that of the CFO and RSSI, while its
variation of (26) is higher than that of the CFO and RSSI.

\begin{figure}[htbp]
\centering
\includegraphics[width=12cm,height=9cm]{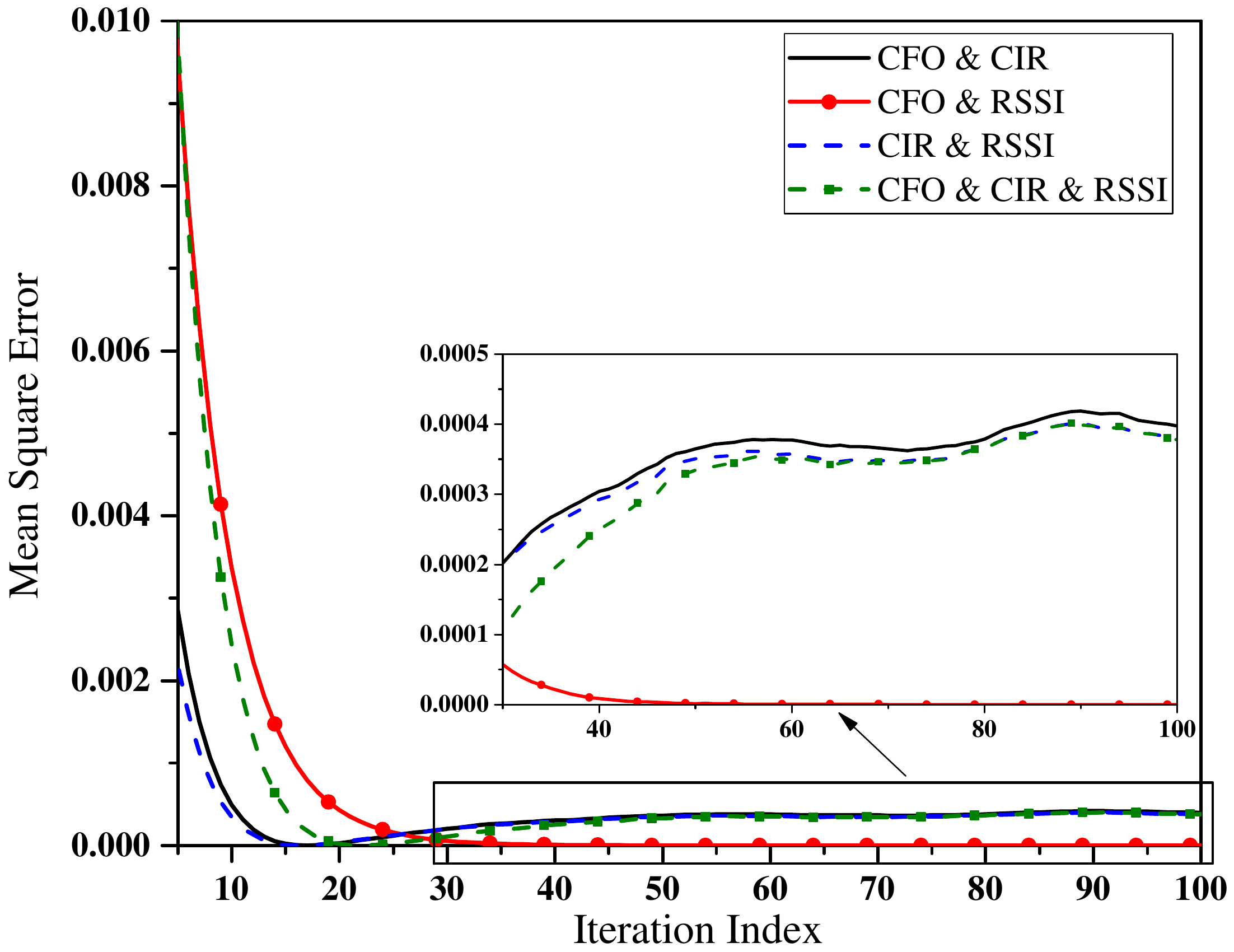}
\caption{\footnotesize{Training performance of our  intelligent authentication process (Algorithm 1) relying on 2 attributes (i.e. CFO $\&$ CIR, CFO $\&$ RSSI, CIR $\&$ RSSI) and 3 attributes (i.e. CFO $\&$ CIR $\&$ RSSI).}}
%\label{fig:fig13}
\end{figure}

Fig. 5 characterizes the training performance of our intelligent authentication process (see Algorithm 1) relying on multiple attributes. We consider four cases, namely the CFO $\&$ CIR, the CFO $\&$ RSSI, the CIR $\&$ RSSI, and finally the CFO $\&$ CIR $\&$ RSSI scenarios. We can observe from Fig. 5 that our intelligent authentication process relying on the CFO $\&$ RSSI pair has the worst training performance before 25 iterations, while that relying on the CIR $\&$ RSSI pair has the lowest mean square error. The reason for this trend is that the mean square error of our intelligent authentication process relying on the CIR is lower than that of the CFO and RSSI before 25 iterations seen in Fig. 4, which adversely affects the training performance in this communication scenario. Additionally, the mean square error of our intelligent authentication process relying on the CFO $\&$ RSSI pair is the lowest after 30 iterations, because both the CFO and RSSI are more reliable than the CIR in the authentication process. Furthermore, it is also shown in Fig. 5 that the training performance of our intelligent authentication process relying on the CFO $\&$ CIR $\&$ RSSI triplet is worse than that of  the CFO $\&$ RSSI pair after 30 iterations, while it is better than that of the CFO $\&$ CIR pair and  CIR $\&$ RSSI pair. This is because  the training performance of our intelligent authentication process depends on both the specific attributes and on the number of physical layer attributes.

\begin{figure}[htbp]
\centering
\includegraphics[width=12cm,height=9cm]{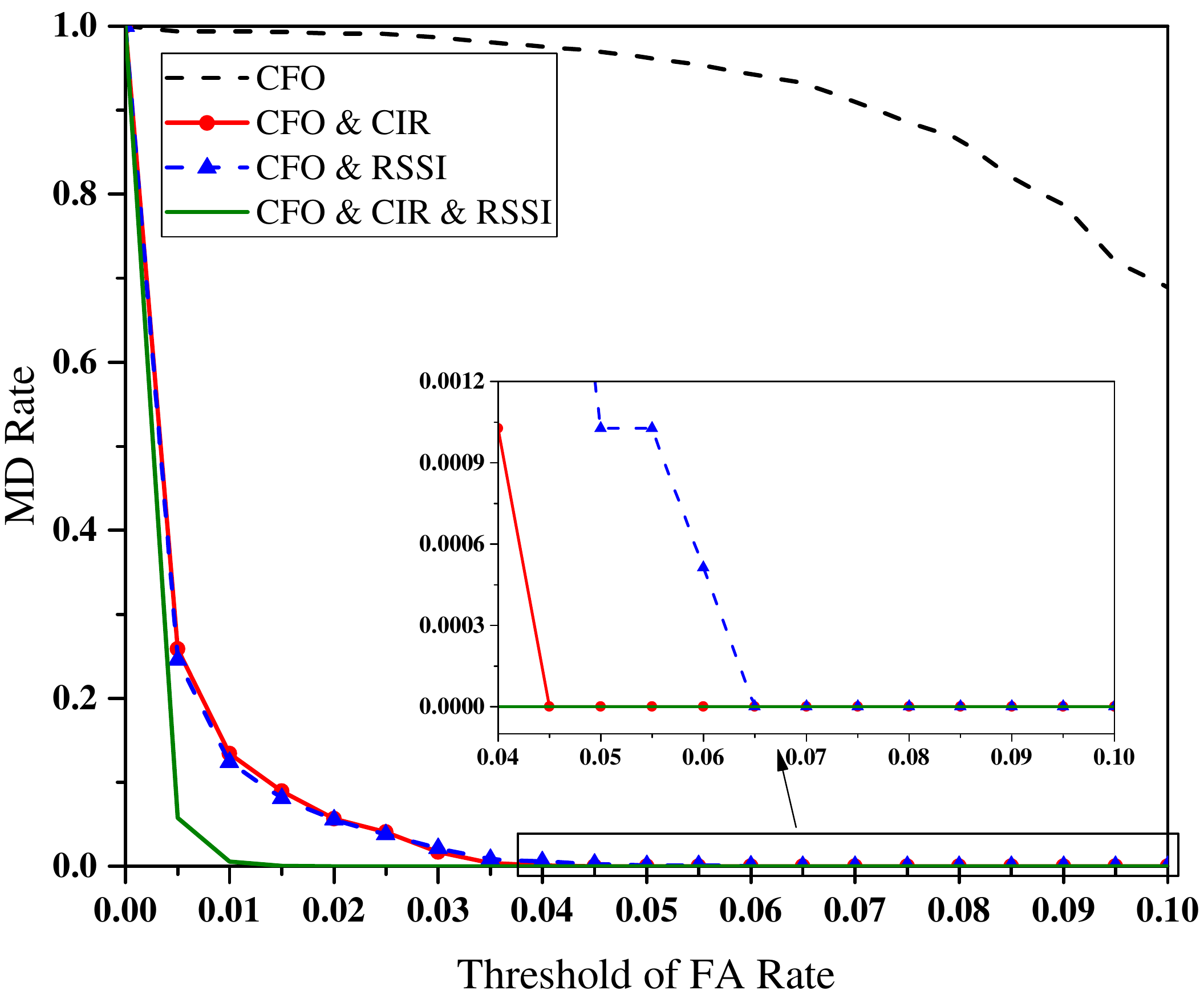}
\caption{\footnotesize{Authentication performance of our intelligent authentication process relying on the CFO, CFO $\&$ CIR, CFO $\&$ RSSI, and CFO $\&$ CIR $\&$ RSSI scenarios. In this case, Eve can intercept and imitate the CFO of Alice.}}
%\label{fig:fig13}
\end{figure}

Fig. 6 considers the case that Eve can intercept and imitate the CFO of Alice, which characterizes the authentication performance of our intelligent authentication process relying on the CFO, CFO $\&$ CIR, CFO $\&$ RSSI, and finally the CFO $\&$ CIR $\&$ RSSI scenarios. In other words, Eve intercepts and impersonates the CFO of Alice to obtain unintended advantages from Bob in this case.
We can observe from Fig. 6 that our intelligent authentication process relying on the CFO $\&$ CIR $\&$ RSSI has the best authentication performance, while that only relying on the CFO performs worst. The reason for this trend is that Bob can better identify the transmitter by using CIR and RSSI, although Eve imitates the CFO of Alice in the  CFO $\&$ CIR $\&$ RSSI scenario. On the other hand, Bob suffers from a high misdetection rate in the CFO scenario, since the CFO of Alice is impersonated by Eve.
It is also shown in Fig. 6 that there is a very small difference between the authentication performance  of our intelligent authentication process relying on the CFO $\&$ CIR pair and that of the CFO $\&$ RSSI pair; and the authentication performances of these two attributes scenarios are better than that of a single-attribute scenario (i.e. CFO). This is because Bob can identify the adversary by using CIR or RSSI in the CFO $\&$ CIR or the CFO $\&$ RSSI scenarios. Therefore,
 the increasing number of physical layer attributes is expected to lead to a higher authentication performance in our intelligent authentication process.

\begin{figure}[htbp]
\centering
\includegraphics[width=12cm,height=9cm]{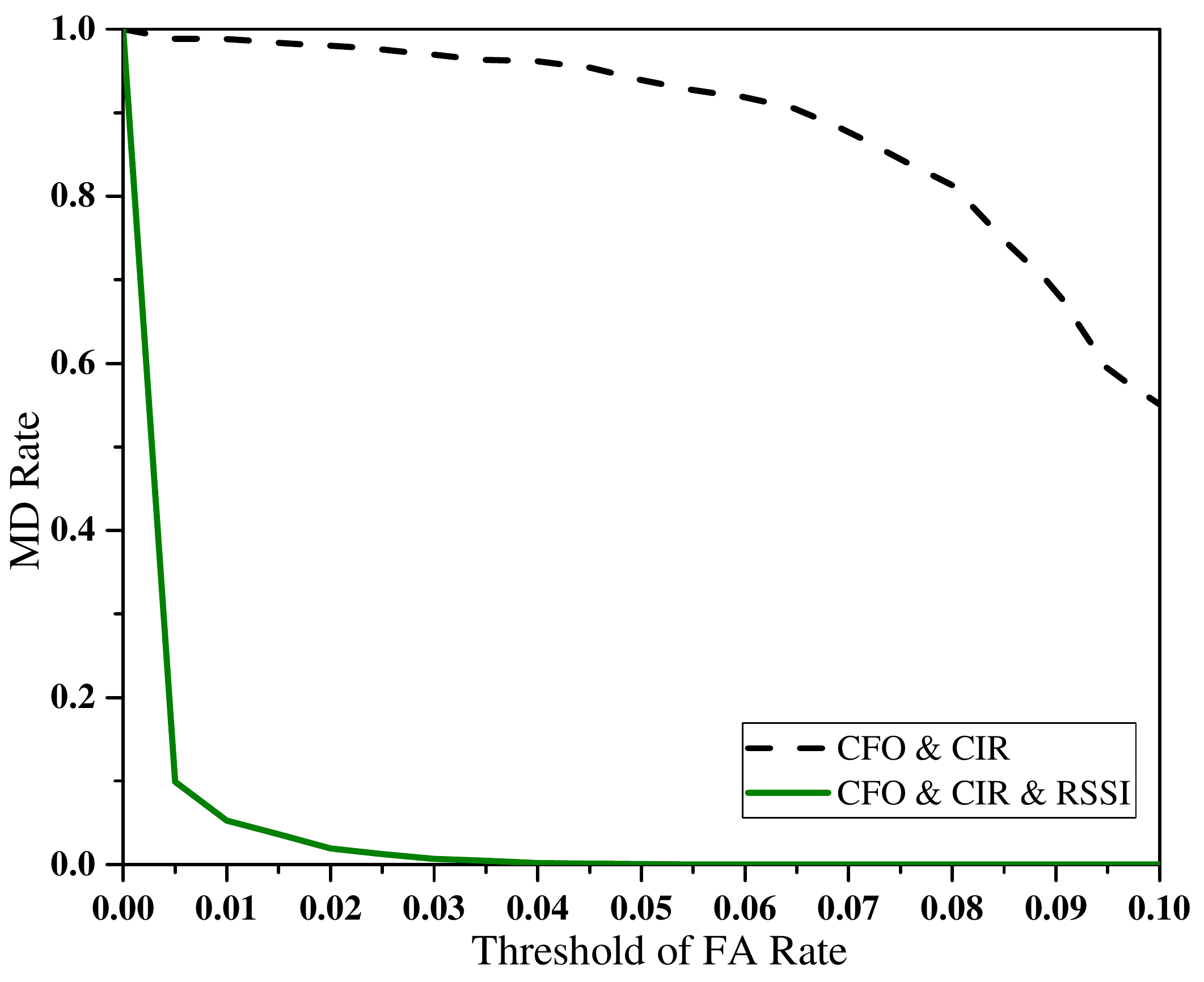}
\caption{\footnotesize{Authentication performance of our intelligent authentication process relying on the CFO $\&$ CIR and CFO $\&$ CIR $\&$ RSSI scenarios. In this case, Eve can intercept and imitate both the CFO and CIR of Alice.}}
%\label{fig:fig13}
\end{figure}

Fig. 7 considers the scenario when Eve can intercept and impersonate both the CFO and CIR of Alice. It is observed from Fig. 7 that the authentication performance of our intelligent authentication process relying on the CFO $\&$ CIR $\&$ RSSI triplet is better than that of the  CFO $\&$ CIR pair. The reason for this trend is that Bob can identify the adversary using the RSSI in the CFO $\&$ CIR $\&$ RSSI scenario, although Eve imitates both the CFO and CIR of Alice. Both Fig. 6 and Fig. 7 confirm that increasing the number of physical layer attributes leads to a better authentication performance, since it is more difficult for an adversary to succeed in
predicting or imitating all the attributes based on the received signal.

\begin{figure}[htbp]
\centering
\includegraphics[width=12cm,height=9cm]{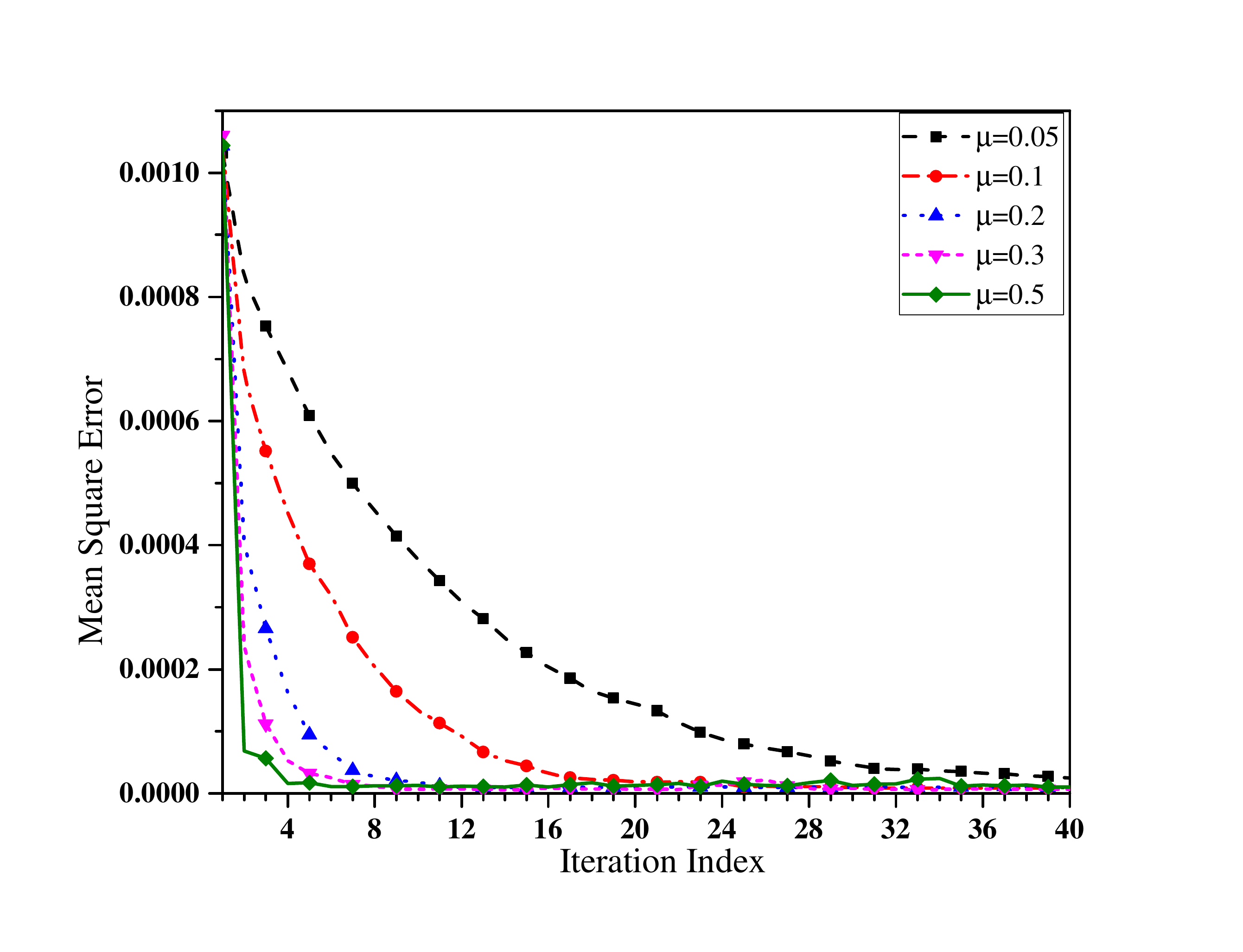}
\caption{\footnotesize{Training performance comparison results of our intelligent authentication process with different step-sizes, i.e. $\mu=0.05$, $\mu=0.1$, $\mu=0.2$, $\mu =0.3$, and $\mu=0.5$.}}
%\label{fig:fig13}
\end{figure}

In Fig. 8, we characterize the training performance of our  intelligent authentication process  (see Algorithm 1) parameterized by the step-sizes of $\mu =0.05$, $\mu =0.1$, $\mu=0.2$, $\mu =0.3$, and $\mu=0.5$. It can be observed from Fig. 8 that our  intelligent authentication process reaches its steady-state value in all cases. We can also see from Fig. 8 that our intelligent authentication process having a higher step-size $\mu$ converges quicker.
In other words, increasing the step-size of learning in a specific range accelerates the convergence. This augments the convergence analysis of Section IV-B.

\begin{figure}[htbp]
\centering
\includegraphics[width=12cm,height=9cm]{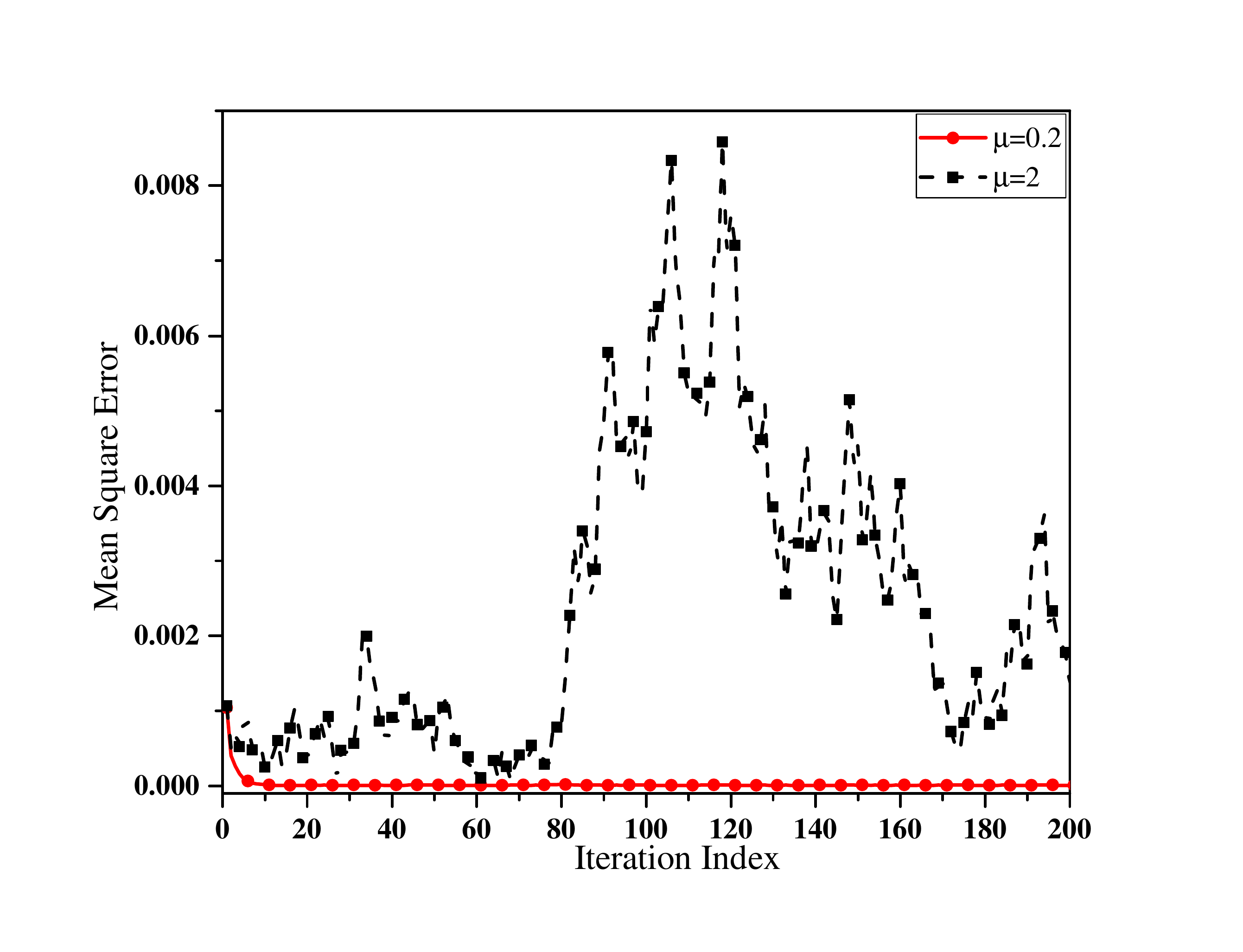}
\caption{\footnotesize{Convergence and divergence of our intelligent authentication process.}}
%\label{fig:fig13}
\end{figure}

Fig. 9 characterizes the mean square error vs. the iteration index for the step-size parameters of $\mu=0.2$ and $\mu=2$. As  discussed before, our authentication process associated with $\mu=0.2$ converges to a steady-state value, while $\mu=2$ diverges. This is because $\mu=2$ is out of the range specified in Theorem 3. Note that the upper bound of the step-size in this case is 1, which can be obtained from (19).

\begin{figure}[htbp]
\centering
\includegraphics[width=12cm,height=9cm]{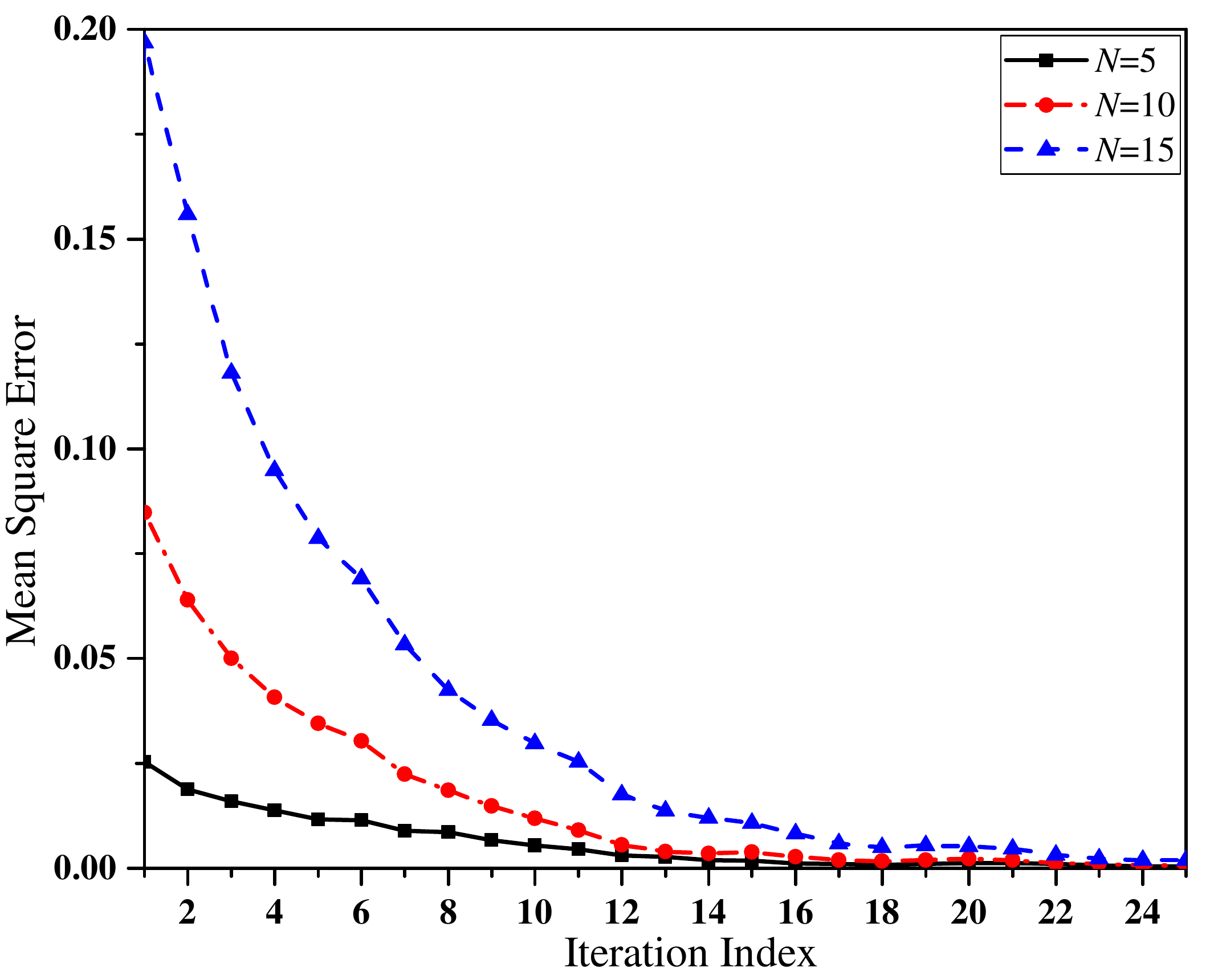}
\caption{\footnotesize{Training performance comparison results of our intelligent authentication process with different numbers of physical layer attributes, i.e., $N=5$, $N=10$ and $N=15$.}}
%\label{fig:fig13}
\end{figure}

Fig. 10 quantifies the influence of the number of physical layer attributes $N$ on the training performance, which shows the mean square error $ E[\| e[l]\|^{2}]$ vs. the iteration index for different numbers of physical layer attributes, namely for $N=5$, $N=10$ and $N=15$. The step-size parameter is set to $\mu=0.1$.
It can be observed that the mean square error $ E[\| e[l]\|^{2}]$ tends to a steady-state value, as the iteration index increases.
Moreover, we can also observe from Fig. 10 that
a larger number of attributes only leads to a slightly slower convergence. Therefore, the explosion of computational complexity
upon increasing the number of physical layer attributes can be readily avoided by our intelligent authentication process. This validates our analysis provided in Section III, and supported by Remark 2, 3, 4.

\begin{figure}[htbp]
\centering
\includegraphics[width=12cm,height=9cm]{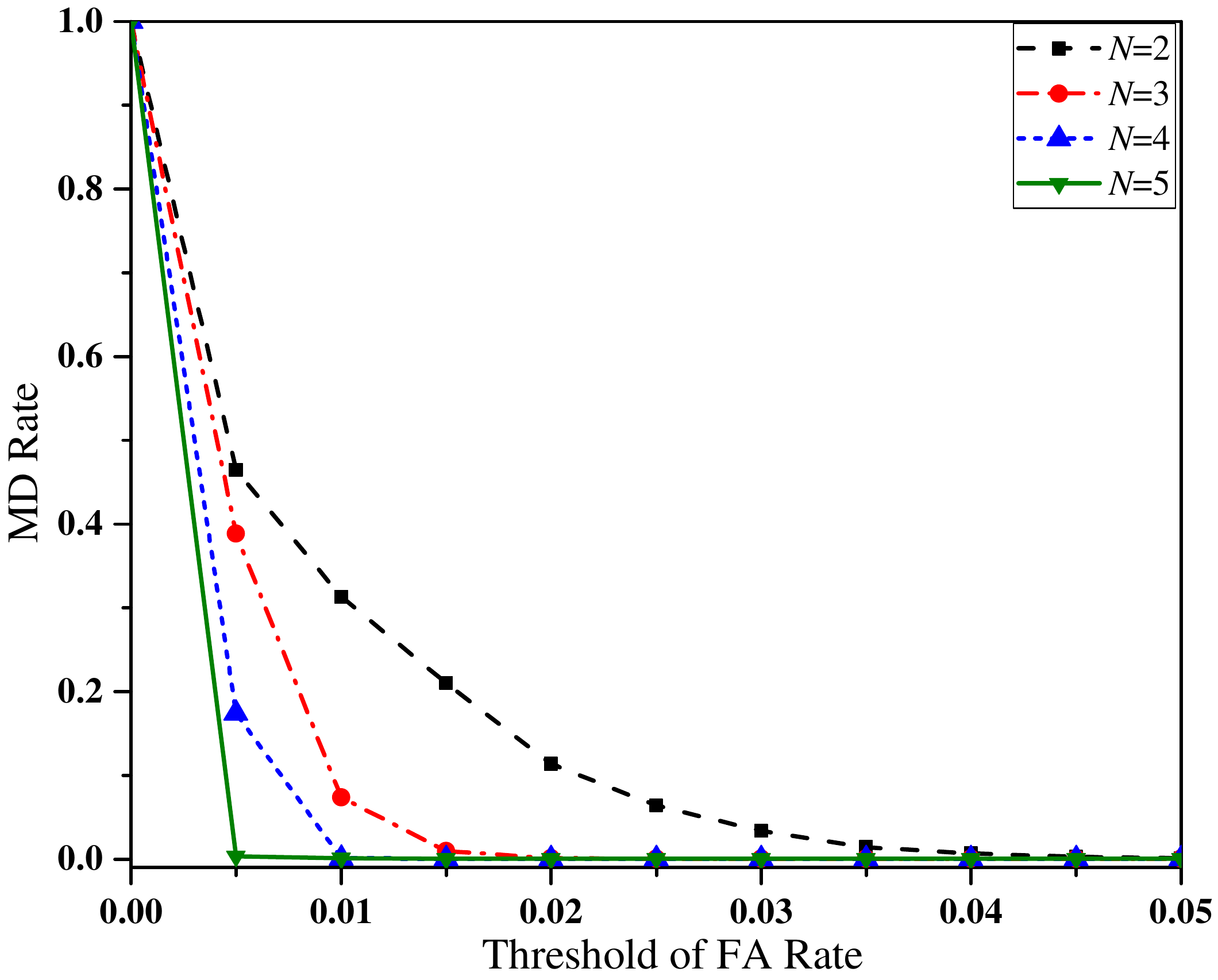}
\caption{\footnotesize{Authentication performance comparison results of our intelligent authentication process with different numbers of physical layer attributes, i.e. $N=2$, $N=3$, $N=4$ and $N=5$.}}
%\label{fig:fig13}
\end{figure}

Fig. 11 characterizes the influence of the number of physical layer attributes $N$ on the authentication performance, which quantifies the MD rate vs. the threshold of FA rate for different numbers of physical layer attributes, namely for $N=2$, $N=3$, $N=4$ and $N=5$. It can be observed that the MD rates are reduced in all cases as the threshold $\delta$ of FA rate increases from 0 to
0.05, because there is an inevitable
 FA-and-MD trade-off.
 One can also observe from Fig. 11 that
a larger number of attributes leads to a more
obvious security performance improvement, without substantially  degrading the convergence performance (see Fig. 10) of our intelligent authentication process. This trend demonstrates the validity of our authentication performance analysis in Section IV-C. In a nutshell, by using more physical layer attributes, our intelligent authentication process achieves a better authentication performance, indicating the presence of a  FA-and-MD trade-off, because we can readily fuse multiple physical layer attributes and control the authentication system to track the variations of multiple attributes. On the same note, the attackers find it more difficult to predict and imitate a larger number of attributes from a received signal.

\begin{figure}[htbp]
\centering
\includegraphics[width=12cm,height=9cm]{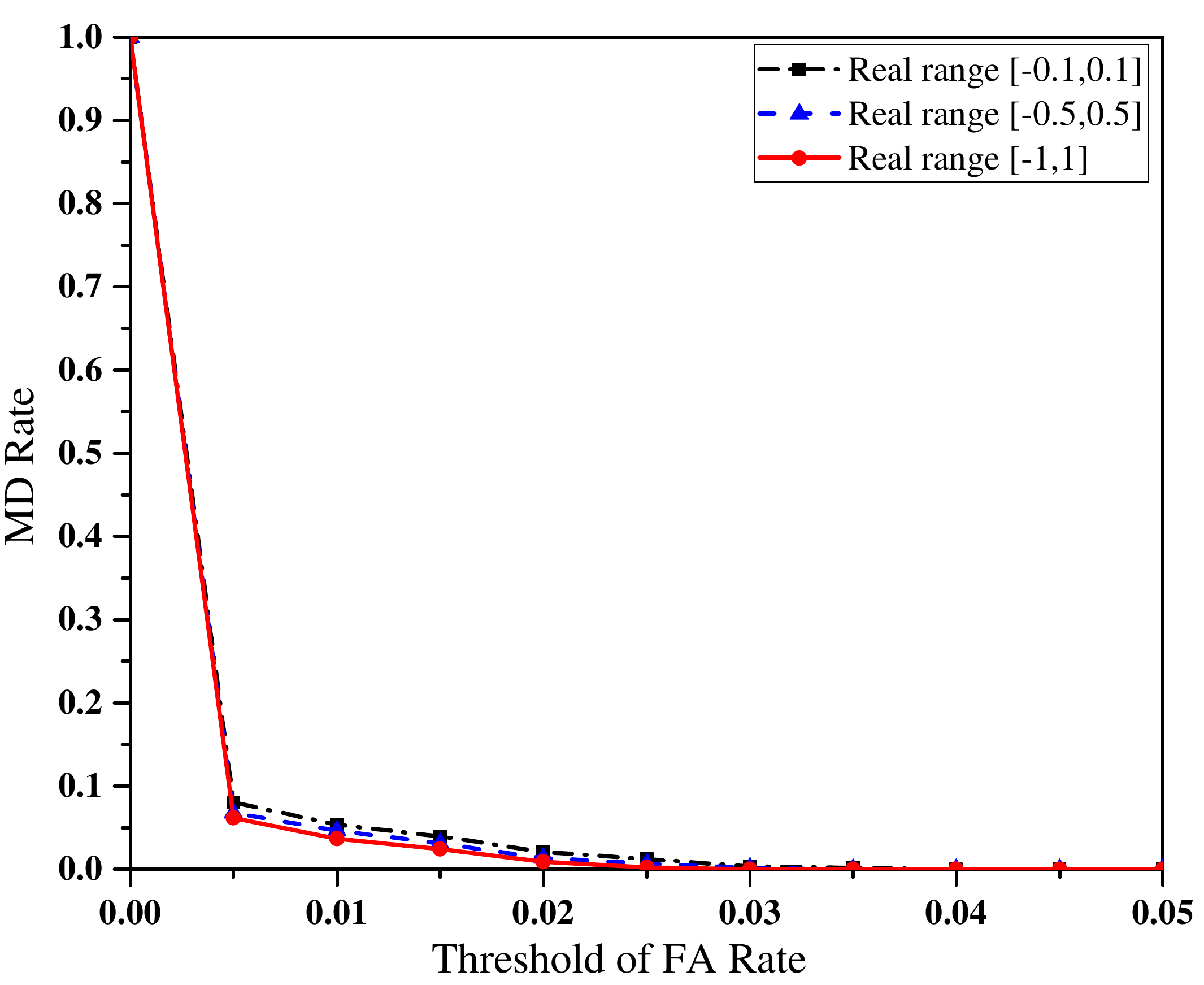}
\caption{\footnotesize{Authentication performance comparison results of our intelligent authentication process with different real ranges in the normalization of Fig. 3 and (5), i.e. $[-0.1,0.1]$, $[-0.5,0.5]$ and $[-1,1]$.}}
\label{fig:fig13}
\end{figure}

Fig. 12 characterizes the false alarm rate vs. the misdetection rate for different real ranges, namely for $[-0.1,0.1]$, $[-0.5,0.5]$ and $[-1,1]$. The real range of $[-0.1,0.1]$ represents the case that  $a_{n}$ and $b_{n}$ of (5) are chosen to be ten times larger than the exact range of attribute $n$ in the normalization of Fig. 3. Therefore, the estimates of attribute $n$ will be scaled to a real range of $[-0.1,0.1]$ instead of the nominal range of $[-1,1]$. Similarly, the real range of $[-0.5,0.5]$ represents the case that we choose $[a_{n},b_{n}]$ twice larger than the exact range of the physical layer attribute $n$. We can observe from Fig. 12 that a mismatch only leads to a small  authentication performance difference, even though we opted for $[a_{n},b_{n}]$ to be ten times larger than the exact range of physical layer attribute $n$.

\begin{figure}[htbp]
\centering
\includegraphics[width=12cm,height=9cm]{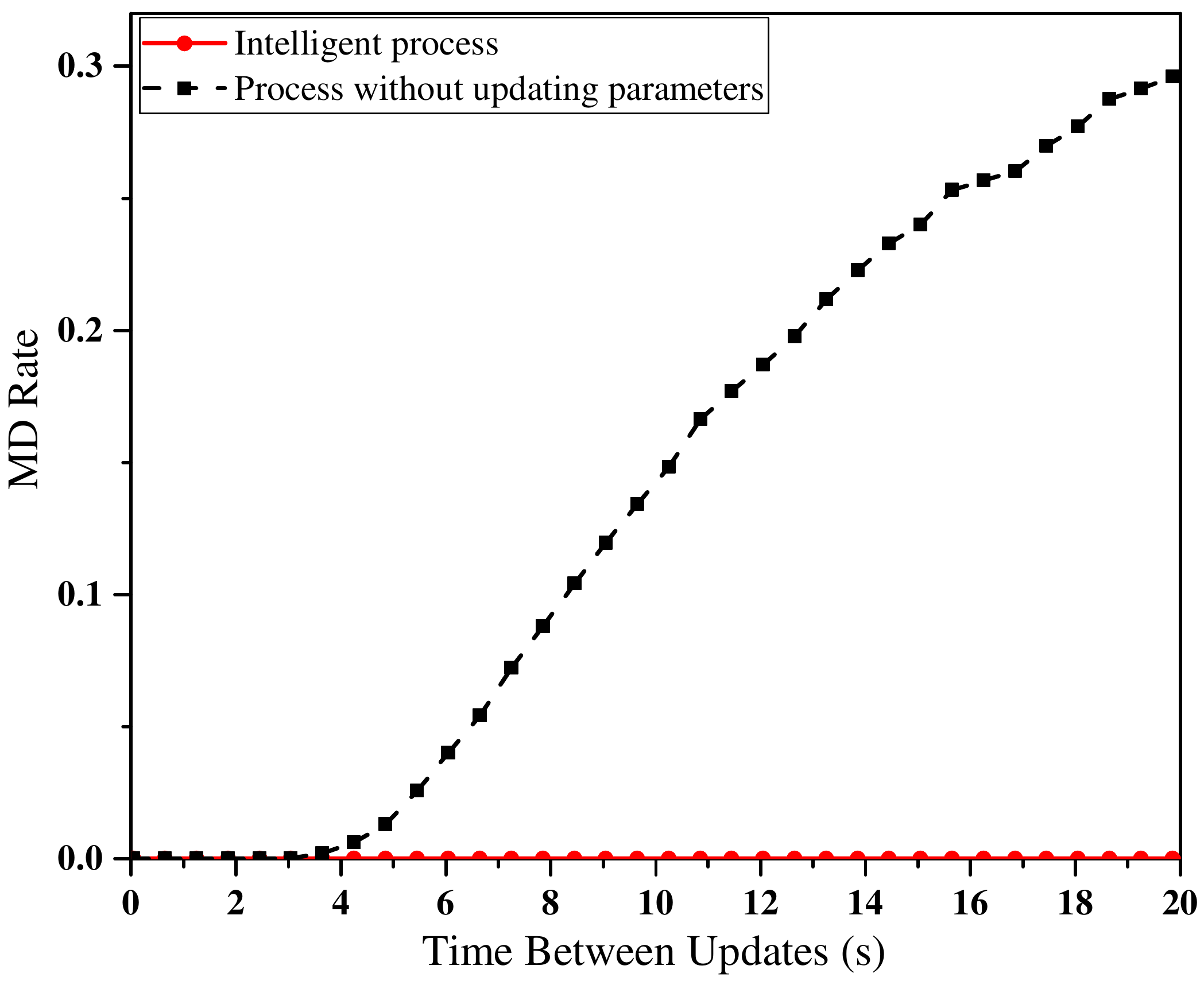}
\caption{\footnotesize{Comparison results between our intelligent process and the process without updating system parameters relying on CFO $\&$ CIR $\&$ RSSI. }}
%\label{fig:fig13}
\end{figure}

In Fig. 13, let us now impose the variations on the CFO, CIR and RSSI for comparing  our intelligent process and the process operating without updating the system parameters. The threshold of the false alarm rate is 0.015.
Then we can observe from Fig. 13 that upon increasing the time between updates, the MD rate of our intelligent process remains robust, tending to $2\times 10^{-5}$, while that of the process operating without updating the system parameters increases dramatically from about $2\times 10^{-5}$ to almost 0.3. This demonstrates that without an adaptive scheme, the authentication performance will be dramatically reduced in time-varying environments.
Therefore,
our intelligent process performs better than the authentication scheme operating without updating the system parameters.

\section{CONCLUSIONS}
In this paper, we proposed an intelligent physical layer authentication technique. A kernel machine-based model was proposed for combining the multiple physical layer attributes and for modelling the authentication as a
 linear system. Through the kernel machine-based multiple attribute fusion model, the number of dimensions of the search-space was reduced from $N$ to 1, and the learning objective was formulated as a convex problem. Therefore, its complexity
was substantially reduced. Then, by conceiving an adaptive authentication process relying on the kernel machine-based multiple attribute fusion model, the process advocated readily accommodated a time-varying environment by discovering and learning this complex dynamic environment. Both the
convergence performance and the authentication performance of our intelligent authentication process
 were theoretically analyzed and numerically validated. The simulation results showed that the authentication performance can be dramatically improved by increasing the number of physical layer attributes exploited by our intelligent authentication process
 without degrading its convergence performance. It was also demonstrated that it has a much better authentication performance in a time-varying environment than its non-adaptive counterpart.
\ifCLASSOPTIONcaptionsoff
  \newpage
\fi

\appendices

~\\

\section{The proof of Theorem 1}
Let
\begin{eqnarray}
J(\bm{w})=\sum _{k=1}^{l}[\widehat{y}_{k}-\bm{w}^{{\rm{T}}}\varphi (\bm{\widetilde{h}}_{k})]^{2}.
\end{eqnarray}
By invoking a step-size parameter $\mu$, the learning rule for the parameter $\bm{w}$
can be derived by using the gradient. The partial derivative of the function $J(\bm{w})$ with respect to $\bm{w}=(w_{1},w_{2},...,w_{l})^{{\rm{T}}}$ is given by
\begin{eqnarray}
\frac{\partial J(\bm{w})}{\partial \bm{w}}=-2 \sum _{k=1}^{l}\varphi (\bm{\widetilde{h}}_{k})[\widehat{y}_{k}-\bm{w}^{{\rm{T}}}\varphi (\bm{\widetilde{h}}_{k})],
\end{eqnarray}
and the instantaneous gradient at iteration $l$ is
\begin{eqnarray}
\frac{\partial J(\bm{w})}{\partial \bm{w}}[l]=- \varphi (\bm{\widetilde{h}}_{l})[\widehat{y}_{l}-\bm{w}[l-1]^{{\rm{T}}}\varphi (\bm{\widetilde{h}}_{l})].
\end{eqnarray}

According to the steepest descent algorithm, we have
\begin{eqnarray}
\bm{w}[l]=\bm{w}[l-1]+ \mu\varphi (\bm{\widetilde{h}}_{l})[\widehat{y}_{l}-\bm{w}[l-1]^{{\rm{T}}}\varphi (\bm{\widetilde{h}}_{l})].
\end{eqnarray}
Since $e[l]$ of (16) can also be expressed as
\begin{eqnarray}
e[l]=\widehat{y}_{l}-\bm{w}[l-1]^{{\rm{T}}}\varphi (\bm{\widetilde{h}}_{l}),
\end{eqnarray}
 the repeated application of (32) through iterations becomes
\begin{eqnarray}\nonumber
\bm{w}[l]=\bm{w}[l-1]+ \mu\varphi (\bm{\widetilde{h}}_{l})e[l]
=\bm{w}[l-2]+ \mu\varphi (\bm{\widetilde{h}}_{l-1})e[l-1]+ \mu\varphi (\bm{\widetilde{h}}_{l})e[l]\\
=\cdots
=\sum _{i=1}^{l}\mu\varphi (\bm{\widetilde{h}}_{i})e[i];~( \bm{w}[0]=0),~l=1,2,...,L.~~~~~~~~~~~~~~~~~~~~
\end{eqnarray}

According to (8), (12) and (13), we can derive the authentication system as
\begin{eqnarray}
f(\bm{\overline{h}})=\sum_{l=1}^{L}\alpha_{l}\kappa(\bm{\widetilde{h}}_{l},\bm{\overline{h}})
=\sum_{l=1}^{L}\alpha_{l}\varphi (\bm{\widetilde{h}}_{l})^{{\rm{T}}}\varphi (\bm{\overline{h}})
=\bm{w}[L]^{{\rm{T}}}\varphi (\bm{\overline{h}})=\sum _{l=1}^{L}\mu e[l]\varphi (\bm{\widetilde{h}}_{l})^{{\rm{T}}}\varphi (\bm{\overline{h}}),
\end{eqnarray}
then we have
\begin{eqnarray}
\alpha_{l}[l]=\mu e[l].
\end{eqnarray}

Therefore, the parameter vector $\bm{\alpha}$ at iteration $l$, i.e., $\bm{\alpha}[l]=(\alpha_{1}[l],\alpha_{2}[l],...,\alpha_{l}[l])^{{\rm{T}}}$, can be updated through (15).~~~~~~~~~~~~~~~~~~~~~~~~~~~~~~~~~~~~~~~~~~~~~~~~~~~~~~~~~~~~~~~~~~~~~~~~~~~~~~~~~~~~~~~~~~$\Box$

\section{The proof of Theorem 2}
According to Theorem 1, the
 authentication system at iteration $l$ can be formulated as
\begin{eqnarray}\nonumber
f(\bm{\overline{h}})[l]=\sum_{k=1}^{l}\alpha_{k}\kappa(\bm{\widetilde{h}}_{k},\bm{\overline{h}})=\mu\sum_{k=1}^{l} e[k]\kappa(\bm{\widetilde{h}}_{k},\bm{\overline{h}})
=\mu\sum_{k=1}^{l-1} e[k]\kappa(\bm{\widetilde{h}}_{k},\bm{\overline{h}})+\mu e[l]\kappa(\bm{\widetilde{h}}_{l},\bm{\overline{h}})\\
=f(\bm{\overline{h}})[l-1]+ \mu e[l]\kappa(\bm{\widetilde{h}}_{l},\bm{\overline{h}}).~~~~~~~~~~~~~~~~~~~~~~~~~~~~~~~~~~~~~~~~~~~~~~~~~~~~~~
\end{eqnarray}
Therefore, the learning rule proposed for adjusting
the authentication system of (12) is expressed as (18). ~~~~~~~~~~~~~~~~~~~~~~~~~~~~~~~~~~~~~~~~~~~~~~~~~~~~~~~~~~~~~~~~~~~~~~~~~~~~~~~~~~~~~~~~~~~~~~~~~~~~$\Box$

\section{The proof of Theorem 3}
A practical convergence criterion is convergence in the mean square error sense, which is formulated as
\begin{eqnarray}
 E[\|e[l]\|^{2}]\rightarrow {\rm{constant}}, {\rm{as}}~l \rightarrow \infty,
\end{eqnarray}
where $E[\cdot]$ represents the expectation of $\cdot$.
It was shown in \cite{24,27} that the least-mean-square criterion based learning is convergent in the mean square, if $\mu$ satisfies
\begin{eqnarray}
0<\mu<\frac{1}{\beta_{max}},
\end{eqnarray}
where $\beta_{max}$ is the largest eigenvalue of the correlation matrix $\bm{\Theta}[L]$ given by
\begin{eqnarray}
\bm{\Theta}[L]=[\varphi (\bm{\widetilde{h}}_{1}), \varphi (\bm{\widetilde{h}}_{2}),...,\varphi (\bm{\widetilde{h}}_{L})]_{N\times L}.
\end{eqnarray}
Since $\beta_{max}< {\rm{tr}}(\bm{\Theta}[L])/L$, where ${\rm{tr}}(\bm{\Theta}[L])$ is the trace of the matrix $\bm{\Theta}[L]$, we have
\begin{eqnarray}
0<\mu<\frac{L}{{\rm{tr}}(\bm{\Theta}[L])}=\frac{L}{\sum_{l=1}^{L}\kappa(\bm{\widetilde{h}}_{l},\bm{\widetilde{h}}_{l})}.
\end{eqnarray}
Therefore, the proposed intelligent authentication process (see Algorithm 1) converges to a steady-state value if the step-size parameter of learning $\mu$ satisfies (19).~~~~~~~~~~~~~~~~~~~~~~~~~~~~~~~~~~~~~~~~~~$\Box$

\section{The proof of Theorem 4}
According to (5), (24), (25), and (26), we can calculate $\bm{\widetilde{h}}_{L}=(\widetilde{h}_{1L},\widetilde{h}_{2L},...,\widetilde{h}_{NL})^{{\rm{T}}}$ in case of $\Phi_{0}$ as
\begin{eqnarray}
\widetilde{h}_{iL}^{\Phi_{0}}=\frac{2}{b_{i}-a_{i}}(\upsilon_{i}(\tau_{L})+\bigtriangleup H_{Ai}^{I}[L-\tau_{L}]
-\bigtriangleup H_{Ai}^{II}[L]-\frac{a_{i}+b_{i}}{2}),
\end{eqnarray}
where $\tau_{L}$ is the time interval between Phase I and Phase II of our physical layer authentication at iteration $L$. Given the distributions of $\bigtriangleup H_{Ai}^{I}$ and $\bigtriangleup H_{Ai}^{II}$ of each physical layer attribute, the probability of density function of $\widetilde{h}_{iL}^{\Phi_{0}}$ can be obtained. Let $Y_{l}=\alpha_{l}\exp(\frac{-\sum_{i=1}^{N}(\widetilde{h}_{il}-\widetilde{h}_{iL}^{\Phi_{0}})^{2}}{2\sigma^{2}})$,
we can calculate the false alarm rate at iteration $L$ as
\begin{eqnarray}\nonumber
P_{FA}=P(|\sum_{l=1}^{L-1}\alpha_{l}\exp(\frac{-\sum_{i=1}^{N}(\widetilde{h}_{il}-\widetilde{h}_{iL}^{\Phi_{0}})^{2}}{2\sigma^{2}})|\leq\nu )
=P(\sum_{l=1}^{L-1}Y_{l}\leq\nu ) -P(\sum_{l=1}^{L-1}Y_{l}<-\nu )~~~~~~\\
=F_{\sum_{l=1}^{L-1}Y_{l}}(\nu)-F_{\sum_{l=1}^{L-1}Y_{l}}(-\nu)
=F_{Y_{1}}\ast F_{Y_{2}}\ast\cdots\ast F_{Y_{L-1}}(\nu)
-F_{Y_{1}}\ast F_{Y_{2}}\ast \cdots\ast F_{Y_{L-1}}(-\nu).~
\end{eqnarray}
Therefore, the false alarm rate expression of our intelligent authentication process at iteration $L$ is shown in (27).~~~~~~~~~~~~~~~~~~~~~~~~~~~~~~~~~~~~~~~~~~~~~~~~~~~~~~~~~~~~~~~~~~~~~~~~~~~~~~~~~~~~~~~~~~~~~~~~~~~~$\Box$

\section{The proof of Theorem 5}
According to (5), (24), and (25), $\bm{\widetilde{h}}_{L}=(\widetilde{h}_{1L},\widetilde{h}_{2L},...,\widetilde{h}_{NL})^{{\rm{T}}}$ in case $\Phi_{1}$ is formulated as
\begin{eqnarray}
\widetilde{h}_{iL}^{\Phi_{1}}=\frac{2}{b_{i}-a_{i}}(\overline{H}_{Ai}^{I}[L-\tau_{L}]-\overline{H}_{Ei}^{II}[L]+\bigtriangleup H_{Ai}^{I}[L-\tau_{L}]
-\bigtriangleup H_{Ei}^{II}[L]-\frac{a_{i}+b_{i}}{2}).
\end{eqnarray}
Given the distributions of $\bigtriangleup H_{Ai}^{I}$ and $\bigtriangleup H_{Ei}^{II}$ of each physical layer attribute, the probability of density function of $\widetilde{h}_{iL}^{\Phi_{1}}$ can be obtained. Upon letting $Z_{l}=\alpha_{l}\exp(\frac{-\sum_{i=1}^{N}(\widetilde{h}_{il}-\widetilde{h}_{iL}^{\Phi_{1}})^{2}}{2\sigma^{2}})$, the misdetection rate at iteration $L$ yields
\begin{eqnarray}\nonumber
P_{MD}
=P(|1-\sum_{l=1}^{L-1}\alpha_{l}\exp(\frac{-\sum_{i=1}^{N}(\widetilde{h}_{il}-\widetilde{h}_{iL}^{\Phi_{1}})^{2}}{2\sigma^{2}})|\leq\nu)~~~~~~~~~~~~~~~~~~~~~\\\nonumber
=P(\sum_{l=1}^{L-1}Z_{l}\leq\nu+1) -P(\sum_{l=1}^{L-1}Z_{l}<1-\nu )~~~~~~~~~~~~~~~~~~~~~~~~~~~~\\
=F_{Z_{1}}\ast F_{Z_{2}}\ast\cdots\ast F_{Z_{L-1}}(\nu+1)
-F_{Z_{1}}\ast F_{Z_{2}}\ast \cdots\ast F_{Z_{L-1}}(1-\nu).
\end{eqnarray}
Therefore, the misdetection rate expression of our intelligent authentication process at iteration  $L$ is given by (28).~~~~~~~~~~~~~~~~~~~~~~~~~~~~~~~~~~~~~~~~~~~~~~~~~~~~~~~~~~~~~~~~~~~~~~~~~~~~~~~~~~~~~~~~~~~~~~~~~~~$\Box$


\begin{thebibliography}{1}
\bibitem{01}
 X. Wang, P. Hao, and L. Hanzo, ``Physical-layer authentication for wireless security enhancement: current challenges and future developments," \emph{IEEE Commun. Mag.}, vol. 54, no. 6, pp. 152-158, 2016.

\bibitem{50}
Y. Zou, J. Zhu, X. Wang, and L. Hanzo, ``A survey on wireless security: technical challenges, recent advances, and future trends," \emph{Proc. IEEE}, vol. 104, no. 9, pp. 1727-1765, 2016.

\bibitem{51}
H. Fang, L. Xu, and X. Wang, ``Coordinated multiple-relay based physical layer security improvement: a single-leader multiple-follower Stackelberg game scheme,"  \emph{IEEE Trans. Inf. Forensics Security}, vol. 13, no. 1, pp. 197-209, 2018.

\bibitem{35}
M. Iwamoto, K. Ohta, and J. Shikata, ``Security formalizations and their relationships for encryption and key agreement in information-theoretic cryptography," \emph{ IEEE Trans. Inf. Theory}, vol. 64, no. 1, pp. 654-685, 2018.

\bibitem{36}
Y. Chen, ``Fully incrementing visual cryptography from a succinct non-monotonic structure," \emph{IEEE Trans. Inf. Forensics Security}, vol. 12, no. 5, pp. 1082-1091, 2017.

\bibitem{47}
Y. Ren, J.-C. Chen, J.-C. Chin, and Y.-C. Tseng, ``Design and analysis of the key management
mechanism in evolved multimedia
broadcast/multicast service," \emph{IEEE Trans. Wireless Commun.}, vol. 15, no. 12, pp.  8463-8476, 2016.




\bibitem{40}
M. Rezaee, P. J. Schreie, M. Guillaud, and B. Clerckx, ``A unified scheme to achieve
the degrees-of-freedom region of the MIMO interference channel with delayed channel state information," \emph{IEEE Trans. Commun.}, vol. 64, no. 3, pp. 1068-1082, 2016.

\bibitem{41}
H. Lohrasbipeydeh, T. A. Gulliver, and H. Amindavar, ``Unknown transmit power RSSD based source
localization with sensor position uncertainty," \emph{IEEE Trans. Commun.}, vol. 63, no. 5, pp. 1784-1797, 2015.

\bibitem{37}
P. Cheng, Z. Chen, F. Hoog, and C. K. Sung, ``Sparse blind carrier-frequency offset estimation for OFDMA uplink," \emph{IEEE Trans. Commun.}, vol. 64, no. 12, pp. 5254-5265, 2016.

\bibitem{38}
O. H. Salim, A. A. Nasir, H. Mehrpouyan, and W. Xiang, ``Multi-relay communications in the presence of phase noise and carrier frequency offsets," \emph{IEEE Trans. Commun.}, vol. 65, no. 1, pp. 79-94, 2017.

\bibitem{39}
A. A. D'Amico, L. Marchetti, M. Morelli, and M. Moretti, ``Frequency estimation in OFDM direct-conversion
receivers using a repeated preamble," \emph{IEEE Trans. Commun.}, vol. 64, no. 3, pp. 1246-1258, 2016.


\bibitem{02}
L. Xiao, L. J. Greenstein, N. B. Mandayam, and W. Trappe, ``Channel-based spoofing detection in frequency-selective rayleigh channels," \emph{IEEE Trans. Wireless Commun.}, vol. 8, no. 12, pp. 5948-5956, 2009.


\bibitem{03}
P. Baracca, N. Laurenti, and S. Tomasin, ``Physical layer authentication over MIMO fading wiretap channels," \emph{IEEE Trans. Wireless Commun.}, vol. 11, no. 7, pp. 2564-2573, 2012.

\bibitem{04}
K. Zeng, K. Govindan, and P. Mohapatra, ``Non-cryptographic authentication and identification in wireless networks," \emph{IEEE Wireless Commun.}, vol. 17, no. 5, pp. 56-62, 2010.

\bibitem{05}
W. Wang, Z. Sun, S. Piao, B. Zhu, and K. Ren, ``Wireless physical-layer identification: modeling and validation," \emph{IEEE Trans. Inf. Forensics Security}, vol. 11, no. 9, pp. 2091-2109, 2016.





\bibitem{07}
A. Ferrante, N. Laurenti, C. Masiero, M. Pavon, and S. Tomasin, ``On the error region for channel estimation-based
physical layer authentication over
Rayleigh fading," \emph{IEEE Trans. Inf. Forensics Security}, vol. 10, no. 5, pp. 941-952, 2015.


\bibitem{08}
W. Wang, Y. Chen, and Q. Zhang, ``Privacy-preserving location authentication
in Wi-Fi networks using fine-grained physical layer signatures," \emph{IEEE Trans. Wireless Commun.}, vol. 15, no. 2, pp. 1218-1225, 2016.


\bibitem{09}
V. Kumar, J. Park, and K. Bian, ``PHY-layer authentication using duobinary
signaling for spectrum enforcement," \emph{IEEE Trans. Inf. Forensics Security}, vol. 11, no. 5, pp. 1027-1038, 2016.


\bibitem{10}
F. Zhu, B. Xiao, J. Liu, and L. Chen, ``Efficient physical-layer unknown tag identification in large-scale RFID systems," \emph{IEEE Trans. Commun.}, vol. 65, no. 1, pp. 283-295, 2016.

\bibitem{11}
G. Caparra, M. Centenaro, N. Laurenti, S. Tomasin, and L. Vangelista, ``Energy-based anchor node selection for IoT physical layer authentication," \emph{in Proc. IEEE International Conference on Communications (ICC)}, pp. 1-6, 2016.

\bibitem{12}
X. Wu, Z. Yang, C. Ling, and X. Xia, ``Artificial-noise-aided physical layer phase
challenge-response authentication for
practical OFDM transmission," \emph{IEEE Trans. Wireless Commun.}, vol. 15, no. 10, pp. 6611-6625, 2016.

\bibitem{06}
W. Hou, X. Wang, J. Chouinard, and A. Refaey, ``Physical layer authentication for mobile systems with time-varying carrier frequency offsets," \emph{IEEE Trans. Commun.}, vol. 62, no. 5, pp. 1658-1667, 2014.

\bibitem{17}
X. Wang, F. J. Liu, D. Fan, H. Tang, and P. C. Mason, ``Continuous physical layer authentication using a novel adaptive OFDM system," \emph{in Proc. IEEE International Conference on Communications (ICC)}, 2011.


\bibitem{13}
X. Duan and X. Wang, ``Authentication handover and privacy protection in 5G HetNets using software-defined networking," \emph{IEEE Commun. Mag.}, vol. 53, no. 4, pp. 28-35, 2015.


\bibitem{14}
J. Liu and  X. Wang, ``Physical layer authentication enhancement using two-dimensional channel quantization," \emph{IEEE Trans. Wireless Commun.}, vol. 15, no. 6, pp. 4171-4182, 2016.


\bibitem{16}
H. Fang, X. Wang, and L. Xu, ``Multiple attributes-based fuzzy physical layer authentication: a hierarchical approach," \emph{Submitted to IEEE Trans. Wireless Commun.}, 2018.


\bibitem{18}
W. Liu, J. C. Principe, and S. Haykin, ``Kernel adaptive filtering: a comprehensive introduction," John Wiley and Sons, pp. 16-98, 2010.

\bibitem{19}
K. Li and J. C. Principe, ``Tranfer learning in adaptive filters: the nearest instance centroid-estimation kernel least-mean-square algorithm," \emph{IEEE Trans. Signal Process.}, vol. 65, no. 24, pp. 6520-6535, 2017.

\bibitem{22}
R. Boloix-Tortosa, J. J. Murillo-Fuentes, I. Santos, and F. Perez-Cruz, ``Widely linear complex-valued kernel methods for regression," \emph{IEEE Trans. Signal Process.}, vol. 65, no. 19, pp. 5240-5248, 2017.

\bibitem{23}
B. Chen, L. Xing, B. Xu, H. Zhao, N. Zheng, and J. C. Principe, ``Kernel risk-sensitive loss: definition, properties, and application to robust adaptive filtering," \emph{IEEE Trans. Signal Process.}, vol. 65, no. 11, pp. 2888-2901, 2017.



\bibitem{62}
J. Liu, P. C. Cosman, and B. D. Rao, ``Robust linear regression via $l_{0}$ regularization," \emph{IEEE Trans. Signal Process.}, vol. 66, no. 3, pp. 698-713, 2018.


\bibitem{63}
G. D. Finlayson, M. Mackiewicz, and A. Hurlbert, ``Color correction using root-polynomial regression," \emph{IEEE Trans. Image Process.}, vol. 24, no. 5, pp. 1460-1470, 2015.

\bibitem{69}
X. Tan, C. Sun, and T. D. Pham, ``Edge-aware filtering with local polynomial
approximation and rectangle-based weighting," \emph{IEEE Trans. Cybern.,} vol. 46, no. 12, pp. 2693-2705, 2016.




\bibitem{70}
L. Yang, L. Zhao, G. Bi, and L. Zhang, ``SAR ground moving target imaging algorithm
based on parametric and dynamic
sparse Bayesian learning," \emph{IEEE Trans. Geosci. Remote Sens.,} vol. 54, no. 4, pp. 2254-2267, 2016.

\bibitem{71}
Y. Li, W. Dong, X. Xie, G. Shi, J.
Wu, and X. Li, ``Image super-resolution with parametric sparse
model learning," \emph{IEEE Trans. Image Process.,} DOI: 10.1109/TIP.2018.2837865, 2018.




\bibitem{61}
S. Wang, Z. Bao, J. S. Culpepper, T. Sellis, and G. Cong, ``Reverse k nearest neighbor
search over trajectories," \emph{IEEE Trans. Knowledge and Data Eng.}, vol. 30, no. 4, pp. 757-771, 2018.

\bibitem{64}
Y.-C. Cheng and Pi-Chung Wang, ``Packet classification using dynamically
generated decision trees," \emph{IEEE Trans. Comput.}, vol. 64, no. 2, pp. 582-586, 2015.

\bibitem{60}
N. Wang, T. Jiang, S. Lv, and L. Xiao, ``Physical-layer authentication based on extreme learning machine," \emph{IEEE Commun. Letters}, vol. 21, no. 7, pp. 1557-1560, 2017.


\bibitem{65}
L. Xiao, Y. Li, G. Han, G. Liu, and W. Zhuang, ``PHY-layer spoofing
detection with reinforcement learning in wireless networks," \emph{IEEE Trans.
Veh. Technol.}, vol. 65, no. 12, pp. 10037-10047, 2016.

\bibitem{67}
P. Abouzar, D. G. Michelson, and M. Hamdi, ``RSSI-based distributed self-localization for wireless sensor networks used in precision agriculture,"  \emph{IEEE Trans. Wireless Commun.}, vol. 15, no. 10, pp. 6638-6650, 2016.



\bibitem{24}
W. Wang, Y. Liang, E. P. Xing, and L. Shen, ``Nonparametric decentralized detection and sparse sensor selection via weighted kernel," \emph{IEEE Trans. Signal Process.}, vol. 64, no. 2, pp. 306-321, 2016.

\bibitem{25}
B. Scholkopf and A. Smola, ``Learning With Kernels," Cambridge, MA, USA: MIT Press, 2002.


\bibitem{28}
W. Hrdle, ``Applied nonparametric regression," Cambridge, UK: Cambridge University Press, 1992.

\bibitem{27}
S. Haykin, ``Adaptive filter theory, 4th edition," Upper Saddle River, NJ: Prentice Hall, 2002.

\bibitem{56}
Y. Shi and M. A. Jensen, ``Improved radiometric identification of wireless devices using MIMO transmission,"  \emph{IEEE Trans. Inf. Forensics Security}, vol. 6, no. 4, pp. 1346-1354, 2011.

\end{thebibliography}
\end{document}